\documentclass[11pt]{article}
  
\usepackage{amsmath}
\usepackage{amssymb}
\usepackage{graphicx}
\usepackage{subfig}
\usepackage{fancyhdr}

\fancypagestyle{plain}{%
\fancyhf{}%
\fancyhead[R]{\it To appear in JCAP}%

}

\newcommand{\eqn}[1]{Eqn. \eqref{#1}}
\newcommand{\eqns}[1]{Eqns. \eqref{#1}}
\newcommand{\Cite}[1]{Ref. \cite{#1}}
\newcommand{\Cites}[1]{Refs. \cite{#1}}
\newcommand{\fig}[1]{Fig. \ref{#1}}
\newcommand{\tab}[1]{Table \ref{#1}}
\newcommand{\ph}[1]{\phantom{#1}}

\newcommand{\ti}[1]{{\tilde #1}}

\begin{document}

\title{Structure Formation, Backreaction and Weak Gravitational Fields}  
\author{{\bf Aseem Paranjape}\footnote{E-mail:
    aseem@tifr.res.in}~~{\bf and}~{\bf T. P. Singh}
\footnote{E-mail: tpsingh@tifr.res.in},\newline\vspace{.1in}\\
{\it Tata Institute of Fundamental Research,}\\
{\it Homi Bhabha Road,}\\ {\it Mumbai 400005, INDIA.}}
\date{}

\maketitle

\begin{abstract}
\noindent There is an ongoing debate in the literature as to whether
the effects of averaging out inhomogeneities (``backreaction'') in 
Cosmology can be large enough to account for the acceleration of the 
scale factor in the FLRW models. In particular, some simple models of 
structure formation studied in the literature seem to indicate that
this is indeed possible, and it has also been suggested that the
perturbed FLRW framework is no longer a good approximation during
structure formation, when the density contrast becomes nonlinear.
In this work we attempt to clarify the situation to some extent, using
a fully relativistic model of pressureless spherical collapse. We find
that whereas averaging during structure formation can lead to
acceleration via a selective choice of averaging  domains, the
acceleration is not present when more generic domains are used for
averaging. Further, we show that for most of the duration of the
collapse, matter velocities remain small, and the perturbed FLRW form
of the metric can be explicitly recovered, in the structure formation
phase. We also discuss the fact that the magnitude of the average
effects of inhomogeneities depends on the scale of averaging, and
while it may not be completely negligible on intermediate scales, it
is expected to remain small when averaging on suitably large scales.
\end{abstract}

\section{Introduction}
Understanding the origin of Dark Energy is undoubtedly one of the most
important problems in cosmology today. Recently, several workers in
the field (including the present authors)
\cite{ras,avg,kolb-notari,wilt}, have supported the hypothesis that
the acceleration of the scale factor in the standard
Friedmann-Lema\^itre-Robertson-Walker (FLRW) cosmologies, might be
attributed to the effects of averaging over inhomogeneities 
in the real Universe. It is known that explicit averaging of the
Einstein equations leads to non-trivial corrections
\cite{ellis,buch,zala}. These corrections or ``backreaction'', it is 
argued, can mimic a Dark Energy term in the Cosmological
equations. Specifically, some simple models have been presented which 
demonstrate how such an effect may arise \cite{ras}.  

On the other hand, it has also been argued that as long as
the standard perturbed FLRW picture for the metric of the Universe is
a good approximation, any effects of averaging over the inhomogeneous
perturbations must remain small (see,
e.g. \Cites{wald,wett,behr,flan}). The counter given to this argument
is that the perturbed FLRW framework is  expected to break down around
the time of structure formation \cite{ras}. That $N$-body simulations
do not demonstrate any such breakdown is attributed to the fact that
these simulations always work in the Newtonian limit of General
Relativity (GR), usually employing periodic boundary conditions, and
that the backreaction due to averaging under such conditions can be
shown to vanish \cite{buch2}. Further, since the evolution of the
background scale factor is fixed in the simulations, backreaction
effects would not be visible even if present.

This situation clearly demands some clarification. One expects that as
long as gravitational fields are weak, the perturbed FLRW framework
should work well. Is the weak field approximation then actually
breaking down during structure formation? If so, then clearly a
radical rethink of the approach to Cosmology is in order. If not, on
the other hand, then one needs to understand how simple models such as
the one presented by R\"as\"anen \cite{ras} can achieve an
acceleration upon averaging during structure formation.
 
In this paper we will attempt to address these issues. We
employ a simple but fully relativistic model of spherical collapse,
\emph{with appropriate initial conditions imposed}. We will show that
the effect observed by R\"as\"anen can actually be attributed to a
very selective choice of averaging domain, and that in fact the effect
is not present when a more generic averaging domain is
chosen. However, we do find that the effect of averaging
inhomogeneities may not be completely negligible. In particular we see
$\sim10\%$ deviations of the effective deceleration parameter to be
defined below, from the expected FLRW value, on the scale over which
we define our averaged quantities. This is in line with the
findings of Li and Schwarz \cite{schwarz,schwarz2} in the context of
averaging of perturbative inhomogeneities. To support these results,
we will show that at any stage of the collapse, \emph{if matter
  velocities are small}, then the weak field conditions hold and one
can explicitly recover the perturbed FLRW form of the metric. We will
see that matter velocities do, in fact, remain small for most of the
duration of the collapse. We will also present an argument explaining
the origin of the $\sim10\%$ effect from perturbative metric
inhomogeneities on the scales at hand, and argue that in the real
universe, the averaging scale is expected to be large enough for these
effects to be much smaller. 

We have organised the paper as follows. In section 2 we will set up
our model for the spherical collapse, and compare its results with
those obtained by R\"as\"anen. In section 3 we will show how the
perturbed FLRW metric is recovered, and we will conclude in section 4
with a discussion of further tests of the backreaction argument.

\section{Spherical Collapse : Setting up the model}

Before we describe the model we use, it will help to recall the model
used by R\"as\"anen \cite{ras}, which is what we wish to compare
with. R\"as\"anen's model can be summarized as follows : one considers
two disjoint regions, one overdense and the other completely empty,
each evolving according to the FLRW evolution equations. (The
embedding of these regions in an FLRW background, and the behaviour of
the region \emph{between} these two regions, is not considered.) The scale
factor in the overdense region therefore behaves as
$a_1\propto(1-\cos{u})$ with $t\propto (u-\sin{u})$, and the scale
factor in the empty region behaves as $a_2\propto t$. It is then
straightforward to show that if one defines a volume averaged scale
factor by $a^3\equiv a_1^3+a_2^3$, then the effective deceleration
parameter given by $q\equiv-(\ddot a a)/{\dot a}^2$ becomes negative
(indicating acceleration) around the time that the overdense region
turns around and starts collapsing. In what follows, we will attempt
to reconstruct such a situation more rigorously, with appropriate
matching and initial conditions taken care of.

\subsection{The LTB solution}
We will employ a model containing spherically symmetric pressureless
``dust'', with the energy-momentum tensor given by 
\begin{equation}
T_{ab} = diag (\rho(t,r),0,0,0)\,.
\label{2eq1}
\end{equation}
The line element of interest is hence the Lema\^itre-Tolman-Bondi
(LTB) metric \cite{LTB} given in synchronous and comoving coordinates,
by 
\begin{equation}
ds^2 = -dt^2 + \frac{R^{\prime2}dr^2}{1-k(r)r^2} + R^2d\Omega^2 \,, 
\label{2eq2}
\end{equation}
where $R(t,r)$ is the area radius of a shell of matter labelled by the
comoving radius $r$, and $k(r)$ is a function of integration to be
determined by the initial conditions. (A prime and a dot refer to
derivatives with respect to $r$ and $t$ respectively. We set the
speed of light $c=1$, except when displaying numerical results where
we will explicitly account for factors of $c$.) The equations of
motion are 
\begin{align}
&{\dot R}^2 = \frac{2GM(r)}{R} - k(r)r^2 \,, 
\label{2eq3}\\
&\rho(t,r) = \frac{M^\prime(r)}{4\pi R^2R^\prime}\,,
\label{2eq4}
\end{align}
where $M(r)$ is another function of integration which can be
determined from the initial density profile once a scaling freedom in
the coordinate $r$ is fixed. The spherical collapse model has been
extensively studied in the literature in the context of structure
formation \cite{sphcoll}. However, to the best of our knowledge, the
issue of whether or not the collapse situation can be recast as
perturbed FLRW, has not been discussed. 

The solutions to \eqn{2eq3} depend on the sign of the function $k(r)$,
and can be given in parametric form as 
\begin{subequations}
\begin{equation}
R(r,t)=\bigg(\frac{9GM(r)}{2}\bigg)^{1/3}
\left(t-t_s(r)\right)^{2/3},~~~~\text{for}~k(r)=0\,,
\label{2eq5a}
\end{equation}
\begin{equation}
R=\frac{GM(r)}{-k(r)r^2}\left(\cosh\eta -1\right)~~~
;~~~t-t_s(r)=\frac{GM(r)}{\left(-k(r)r^2\right)^{3/2}}\left(\sinh\eta  
- \eta\right)~,~0\leq\eta<\infty~,~~~\text{for}~k(r)<0\,,
\label{2eq5b}
\end{equation}
\begin{equation}
R=\frac{GM(r)}{k(r)r^2}\left(1-\cos\eta\right)~~~;~~~
t-t_s(r)=\frac{GM(r)}{\left(k(r)r^2\right)^{3/2}}
\left(\eta -\sin\eta\right)~,~0\leq\eta\leq 2\pi,~~~
\text{for}~k(r)>0\,.
\label{2eq5c} 
\end{equation}
\end{subequations}
The function $t_s(r)$ appearing above is completely determined once
$k(r)$ and $M(r)$ are known and a scaling choice for $r$ is
made. Since we will always start the evolution at $t=t_i>t_s(r)$, this
function will be of little physical relevance, except to ensure
consistency at $t=t_i$.

\subsection{Initial conditions}
While choosing the initial density, velocity and coordinate scaling
profiles, we make the important assumption that at initial
time, a well-defined \emph{global} background FLRW solution can be
identified, with scale factor $a(t)$, Hubble parameter $H(t)$ and
density $\bar\rho(t)$. This is reasonable since the CMB data (combined
with the Copernican principle) assure us that inhomogeneities at the
last scattering epoch were at the level of 10 parts per million. This
assumption plays a crucial role in deciding which regions are
overdense and will eventually collapse, and which regions will keep
expanding (see also the first paper in \Cite{wilt} for a discussion on
this point).
\begin{itemize}
\item {\bf Initial density profile} $\rho(t_i,r)$ : \\
The initial density is chosen to be
\begin{equation}
\rho(t_i,r)=\bar\rho_{i}\left\{
\begin{array}{l}
(1+\delta_\ast),~~~~r<r_\ast\\
(1-\delta_v),~~~r_\ast<r<r_v\\
1,~~~~~~~~~~~~r>r_v\,,
\end{array}\right .
\label{2eq6}
\end{equation}
where $\bar\rho_{i}=\bar\rho(t_i)$. Initially, the region $r<r_\ast$
is assumed to contain a tiny overdensity and the region
$r_\ast<r<r_v$, an underdensity. In other words, 
\begin{equation}
0 < \delta_\ast,\delta_v \ll 1\,.
\label{2eq7}
\end{equation}
The discontinuities in the initial density profile can be smoothed out
by replacing the step functions appropriately. We will not do this
here, since the step functions make calculations very simple. This is
not expected to affect the qualitative features of our final results. 
\item {\bf Initial conditions on scaling and velocities} : \\
We match the initial velocity and coordinate scaling to the global
background solution, by requiring 
\begin{align}
R(t_i,r) &= a_i r\,,
\label{2eq8} \\
\dot R(t_i,r) &= a_i H_i r \,,
\label{2eq9}
\end{align}
with $a_i$ and $H_i$ denoting the initial values of the scale factor
and Hubble parameter respectively of the global background. This
amounts to setting the initial velocities to match the Hubble flow,
ignoring initial peculiar velocities. This is only a convenient choice
and the introduction of initial peculiar velocities is not expected to
modify our final results qualitatively. 
\end{itemize}
For the FLRW background we consider an Einstein-deSitter (EdS)
solution with scale factor and Hubble parameter given by 
\begin{align}
&a(t) = (t/t_0)^{2/3} ~~;~~ t_0 = 2/(3H_0) \,,
\label{2eq10} \\
&H(t) \equiv \dot a/a = 2/(3t)\,,
\label{2eq11}
\end{align}
with $t_0$ denoting the present epoch. $a_i$ fixes the initial time as 
\begin{equation}
t_i = 2/(3H_0) a_i^{3/2}\,.
\label{2eq12}
\end{equation}
We will always use $a_i=10^{-3}$, so that the initial conditions are
being set around the CMB last scattering epoch; in general $a_i$ must
be treated as one of the parameters in the problem. The initial EdS
background density is given in terms of $H_0$ and $a_i$ as
\begin{equation}
\bar\rho_{i}= \frac{3}{8\pi G}H_0^2a_i^{-3}\,.
\label{2eq13}
\end{equation}
\subsection{Mass function $M(r)$ and curvature function $k(r)$}
We now have enough information to fix $M(r)$ and $k(r)$. Using
\eqn{2eq4} at initial time together with the scaling in \eqn{2eq8}
gives us
\begin{equation}
GM(r)=\frac{1}{2}H_0^2r^3\left\{ 
\begin{array}{l}
1 + \delta_\ast,~~~~0<r<r_\ast\\
1 + \delta_v\left( \left(r_c/r\right)^3 - 1 \right),
~~~r_\ast<r<r_v\\    
1+(\delta_v/r^3)\left(r_c^3 - r_v^3\right),~~r>r_v\,, 
\end{array}\right .
\label{2eq14}
\end{equation}
where we have defined a ``critical'' radius $r_c$ by the equation
\begin{equation}
\left(\frac{r_c}{r_\ast}\right)^3 = 1 + \frac{\delta_\ast}{\delta_v} 
\,. 
\label{2eq15}
\end{equation}
The significance of $r_c$ will become apparent shortly.
Using the initial conditions \eqns{2eq8} and \eqref{2eq9} in the
evolution equation \eqref{2eq3} at initial time, gives
\begin{equation}
k(r) r^2 = \frac{2GM(r)}{a_i r} - a_i^2 H_i^2 r^2\,,
\label{2eq16}
\end{equation}
with $H_i^2 = H_0^2a_i^{-3}$, and hence
\begin{equation}
k(r)=\frac{H_0^2}{a_i}\left\{ 
\begin{array}{l}
\delta_\ast,~~~~r<r_\ast\\
\delta_v\left(\left(r_c/r\right)^3-1\right), ~~~ r_\ast<r<r_v\\    
(\delta_v/r^3)\left(r_c^3 - r_v^3\right), ~~ r>r_v\,.  
\end{array}\right .
\label{2eq17}
\end{equation}
The significance of $r_c$ is now clarified. Since
$\delta_\ast,\delta_v>0$, we have $r_c>r_\ast$ by definition
(\eqn{2eq15}). The following possibilities arise :
\begin{itemize}
\item If $r_c>r_v$, then $k(r)>0$ for all $r$, and every shell will 
ultimately collapse, including the ``void'' region $r_\ast<r<r_v$. 
\item If $r_c<r_v$, then $k(r)>0$ for $r<r_c$ and changes sign at
  $r=r_c$. Hence, the region $r_\ast<r<r_c$ will collapse even
  though it is underdense, while the region $r>r_c$ will expand
  forever.  
\item If $r_c=r_v$, then the ``void'' exactly compensates for the
  overdensity, and the universe is exactly EdS for $r>r_v$. [$GM(r) = 
  (1/2)H_0^2r^3$ and $k(r)=0$.] Also the ``void'' will eventually
  collapse.  
\end{itemize}
Clearly the most interesting case for us is the one with $r_c<r_v$,
and we will hence make this choice for our model. We realize that the
model as it stands is not a very realistic depiction of the (nearly
spherical) voids we see in our Universe \cite{voids}, since these
voids are seen to be \emph{surrounded} by ``walls'' of
matter. However, our goal is to describe two regions, one of which
collapses while the other expands ever more rapidly, and our model is
capable of doing so while retaining its fully relativistic
character. 

Although we have set up the model for all values of the radial
coordinate $r$, hereon we will concentrate on the region
$0<r<r_v$. One reason is that most of the interesting dynamics takes
place in this region. Another is that the region $r>r_v$ develops
shell-crossing singularities due to the sharp rise in density across
$r=r_v$. A more realistic model would be able to incorporate the
pressures that are expected to build up when a shell-crossing occurs
(\cite{bert}, see also the discussion in section 4 below), but the LTB
model is limited in this respect due to its pressureless character. We
will therefore ignore the region $r>r_v$.

\subsection{The solution in the region $0<r<r_v$}
The region of interest can be split into three parts : Region 1
$=\left\{0<r<r_\ast\right\}$, Region 2 $=\left\{r_\ast<r<r_c\right\}$
and Region 3 $=\left\{r_c<r<r_v\right\}$. The solution in the three
regions is as follows :
\begin{itemize}
\item {\bf Region 1 ($0<r<r_\ast$) :}  
\begin{subequations}
\begin{align}
&R = \frac{1}{2}\left(\frac{a_i}{\delta_\ast}\right) r (1+\delta_\ast)
(1-\cos{u}) \,,
\label{2eq18a}\\
&u-\sin{u} = \frac{2H_0}{1+\delta_\ast}
\left(\frac{\delta_\ast}{a_i}\right)^{3/2} (t-t_i) + (u_i - \sin{u_i})
\,, 
\label{2eq18b}\\
&1 - \cos{u_i} = \frac{2\delta_\ast}{1+\delta_\ast} \,,
\label{2eq18c}\\
&R^2R^\prime = \frac{R^3}{r}\,.
\label{2eq18d}
\end{align}
\label{2eq18}
\end{subequations}
\end{itemize}
For Regions $2$ and $3$, it is convenient to define a function
$\varepsilon(r)$ as
\begin{equation}
\varepsilon(r) \equiv \delta_v \left( \left(\frac{r_c}{r}\right)^3 - 1
\right) = \frac{a_i}{H_0^2}k(r) \,, ~~~ r_\ast < r < r_v\,.
\label{2eq19}
\end{equation}
\begin{itemize}
\item {\bf Region 2 ($r_\ast<r<r_c$) :}  
\begin{subequations}
\begin{align}
&R = \frac{1}{2}\left(\frac{a_i}{\varepsilon}\right) r (1+\varepsilon) 
(1-\cos{\alpha}) \,,
\label{2eq20a}\\
&\alpha-\sin{\alpha} = \frac{2H_0}{1+\varepsilon}
\left(\frac{\varepsilon}{a_i}\right)^{3/2} (t-t_i) + (\alpha_i -
\sin{\alpha_i}) \,, 
\label{2eq20b}\\
&1 - \cos{\alpha_i(r)} = \frac{2\varepsilon}{1+\varepsilon} \,,
\label{2eq20c}\\
&R^2R^\prime = \frac{R^3}{r} \left(1 -
  \frac{r\varepsilon^\prime}{\varepsilon(1+\varepsilon)}\left\{1 -
  \frac{\varepsilon^{3/2}}{(1-\cos{\alpha})^2}
  \left[H_i(t-t_i)\sin{\alpha}
  \left(\frac{3+\varepsilon}{1+\varepsilon}\right) \right. \right. \right. 
\nonumber\\    
&\ph{R^2R^\prime = \frac{R^3}{r}\left[1 -
    \frac{r\varepsilon^\prime}{\varepsilon(1+\varepsilon)} \right]  }
  \left. \left. \left. +  \,
\frac{4\varepsilon^{1/2}}{(1+\varepsilon)^2}
\left(\frac{\sin{\alpha}}{\sin{\alpha_i}}\right) \right]\right\}  
  \right)    \,.
\label{2eq20d}
\end{align}
\label{2eq20}
\end{subequations}
\item {\bf Region 3 ($r_c<r<r_v$) :}  
\begin{subequations}
\begin{align}
&R = \frac{1}{2}\left(\frac{a_i}{|\varepsilon|}\right) r (1+\varepsilon)  
(\cosh{\eta}-1) \,,
\label{2eq21a}\\
&\sinh{\eta}-\eta = \frac{2H_0}{1+\varepsilon}
\left(\frac{|\varepsilon|}{a_i}\right)^{3/2} (t-t_i) + (\sinh{\eta_i} -
\eta_i) \,,  
\label{2eq21b}\\
&\cosh{\eta_i(r)}-1 = \frac{2|\varepsilon|}{1+\varepsilon} \,,
\label{2eq21c}\\
&R^2R^\prime = \frac{R^3}{r} \left(1 -
  \frac{r\varepsilon^\prime}{\varepsilon(1+\varepsilon)}\left\{1 -
  \frac{|\varepsilon|^{3/2}}{(\cosh{\eta}-1)^2}
  \left[H_i(t-t_i)\sinh{\eta}
  \left(\frac{3+\varepsilon}{1+\varepsilon}\right) \right. \right. \right. 
\nonumber\\    
&\ph{R^2R^\prime = \frac{R^3}{r}\left[1 -
    \frac{r\varepsilon^\prime}{\varepsilon(1+\varepsilon)} \right]  }
  \left. \left. \left. +  \,
\frac{4|\varepsilon|^{1/2}}{(1+\varepsilon)^2}
\left(\frac{\sinh{\eta}}{\sinh{\eta_i}}\right) \right]\right\}  
  \right)    \,.
\label{2eq21d}
\end{align}
\label{2eq21}
\end{subequations}
\end{itemize}
The crossover from Region 1 to Region 2 is discontinuous in $R^\prime$
(but not in $R$) due to our discontinuous choice of initial
density. Smoothing out the density will also smooth out
$R^\prime$. The crossover from Region 2 to Region 3 can be shown to be
smooth, by considering the limits $r\to r_c^-$ and $r\to r_c^+$ or
equivalently $\varepsilon\to0^-$ and $\varepsilon\to0^+$. Note that the
results in \eqns{2eq18}, \eqref{2eq20} and \eqref{2eq21} are exact,
and do not involve any perturbative expansions in $\delta_\ast$ or
$\delta_v$, even though these parameters are small.

\subsection{Behaviour of the model}
Each shell in the inner, homogeneous and overdense Region 1 behaves as
a closed FLRW universe, expanding out to a maximum radius $R_{max}(r)$
given by 
\begin{equation}
R_{max}(r) = \frac{a_i}{\delta_\ast}r(1+\delta_\ast)\,.
\label{2eq22}
\end{equation}
\begin{table}[t]
\centering
\begin{tabular}{|c|c|}\hline
Parameter name & Parameter value \\ [1ex]\hline \hline
$a_i$ & $0.001$ \\ [0.5ex]\hline
$H_0$ & $~~~1/13.59\, {\rm Gyr}^{-1} ~~ (=72 \,{\rm km/s/Mpc})~~~$  
\\ [0.5ex] \hline  
$t_0$ & $2/(3H_0) = 9.06 {\rm Gyr}$ \\ [0.5ex] \hline
$c$ & $306.6\, {\rm Mpc Gyr}^{-1}$\\ [0.5ex] \hline
$\delta_\ast$ & $1.25 a_i(3\pi/4)^{2/3} = 2.21 \times 10^{-3}$
\\ [0.5ex] \hline 
$\delta_v$ & $0.005$ \\ [0.5ex] \hline 
$r_\ast$ & $0.004 c/H_0 = 16.7\, {\rm Mpc}$ \\ [0.5ex] \hline 
$t_{turn}/t_0$ & $0.72$ \\ [0.5ex] \hline
$r_c$ & $r_\ast\left(1+\delta_\ast/\delta_v\right)^{1/3} = 18.8 \,
   {\rm Mpc}$ \\ [0.5ex] \hline 
$r_v$ & $1.25 r_c = 23.5\, {\rm Mpc}$\\ [0.5ex] \hline 
$R(t_0,r_\ast)$ & $6.8\,{\rm Mpc}$\\ [0.5ex] \hline 
$R(t_0,r_v)$ & $33.3\,{\rm Mpc}$\\ [0.5ex] \hline 
\end{tabular}
\caption{\small Values of various parameters used in generating
  plots.}  
\label{tab1}
\end{table}
All the inner shells reach their maximum radius and turn around at the
same time $t_{turn}$ given by
\begin{equation}
t_{turn} = t_i + \frac{1+\delta_\ast}{2H_0}
\left(\frac{a_i}{\delta_\ast}\right)^{3/2}
\left(\pi-(u_i-\sin{u_i})\right) \approx
t_0\left(\frac{3\pi}{4}\right)
\left(\frac{a_i}{\delta_\ast}\right)^{3/2} \,,
\label{2eq23}
\end{equation}
where we have used the smallness of $a_i$ and $\delta_\ast$ to make
the last approximation. By appropriately choosing a value of
$\delta_\ast$, we can arrange for the turnaround of Region 1 to occur
either before or after the present epoch. 

In \tab{tab1} we have listed the parameter values which we will use
frequently in displaying plots. Along with the parameter set
$\{a_i,H_0,\delta_\ast,\delta_v,r_\ast,r_v\}$, we have also listed the
values of the derived quantities $\{r_c,t_i,t_0,t_{turn}\}$ and speed
of light $c$ in units of ${\rm Mpc Gyr}^{-1}$. We have also shown the
values of the present day physical area radius $R(t_0,r)$ at
$r=r_\ast$ and $r=r_v$. The density contrasts are to be understood to
reflect the inhomogeneities in the dark matter density close to last
scattering, and not the inhomogeneities of the baryons which were much
smaller \cite{dodel,LSE-density}. 

\begin{figure}[t]
\centering
\fbox{
\subfloat[]{\includegraphics[width=.4\textwidth]
{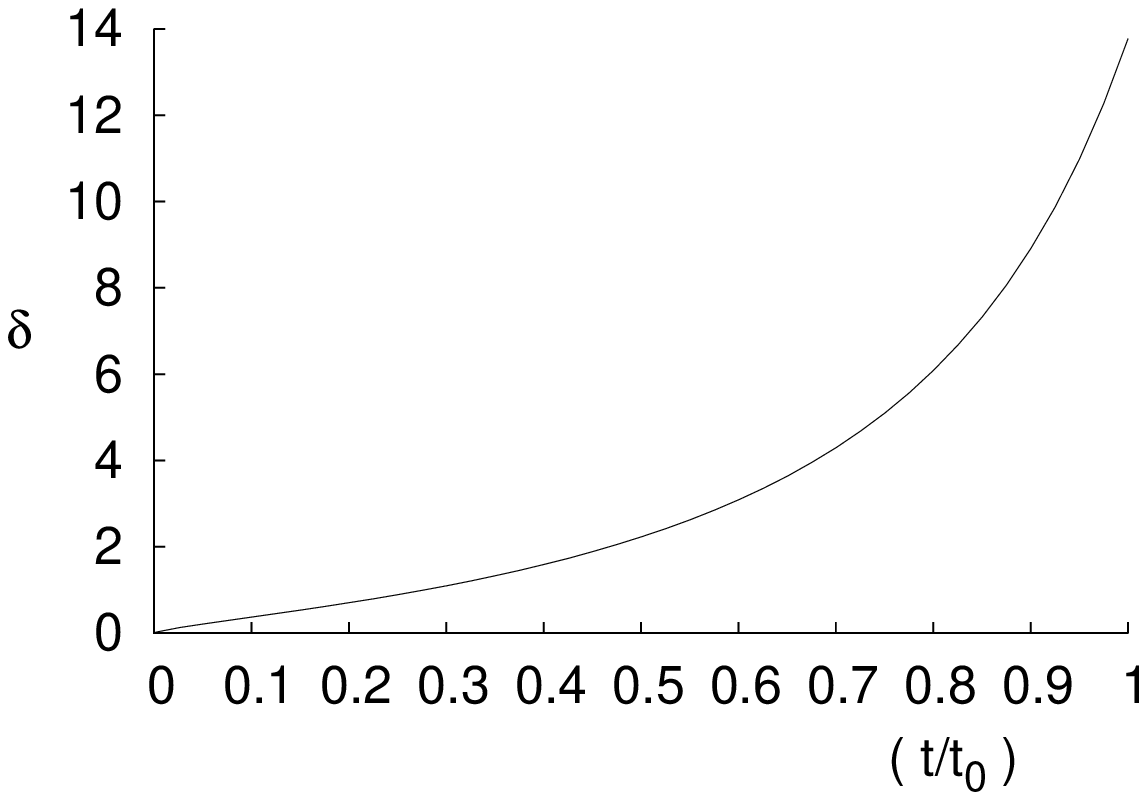}  \label{fig1a}}  
}
\fbox{
\subfloat[]{\includegraphics[width=.4\textwidth]
{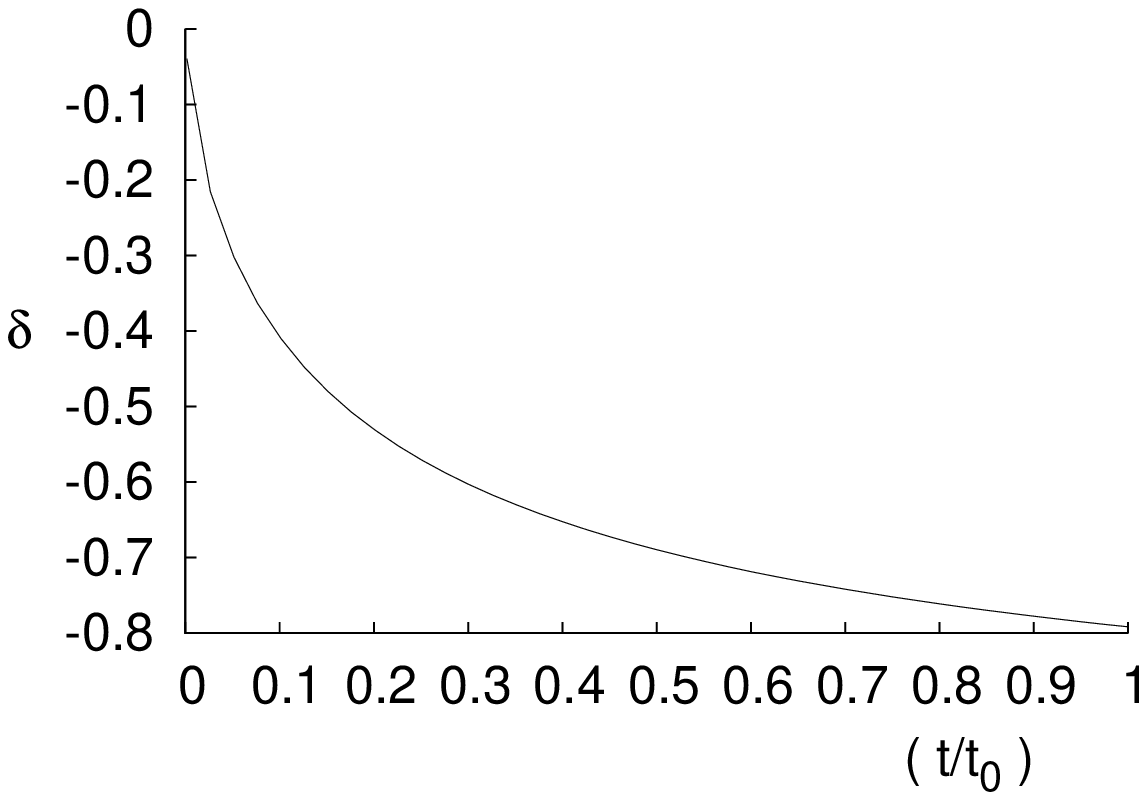}  \label{fig1b}}  
}
\caption{\small The evolution of the density contrast $\delta(t,r)$,
 using parameter values from \tab{tab1} evaluated at (a) $r=r_\ast/2$
 in Region 1 and (b) $r=(r_c + r_v)/2$ in Region 3.}  
\label{fig1}
\end{figure}

In \fig{fig1} we have shown the evolution of the density contrast
$\delta(t,r)$ defined in the usual way by 
\begin{equation}
1+\delta(t,r) = \frac{\rho(t,r)}{\bar\rho(t)}\,,
\label{2eq24}
\end{equation}
for the parameter choices of \tab{tab1}, for which one has
$t_{turn}/t_0 \simeq 0.72$, so that the collapse is well under 
way in Region 1 at the present epoch. The two panels show the contrast
for two representative values of $r$, one in Region 1 and the other in
Region 3.

\begin{figure}[t]
\centering
\fbox{
\subfloat[$t_{turn}= 1.278 t_0$]{\includegraphics[width=.4\textwidth] 
{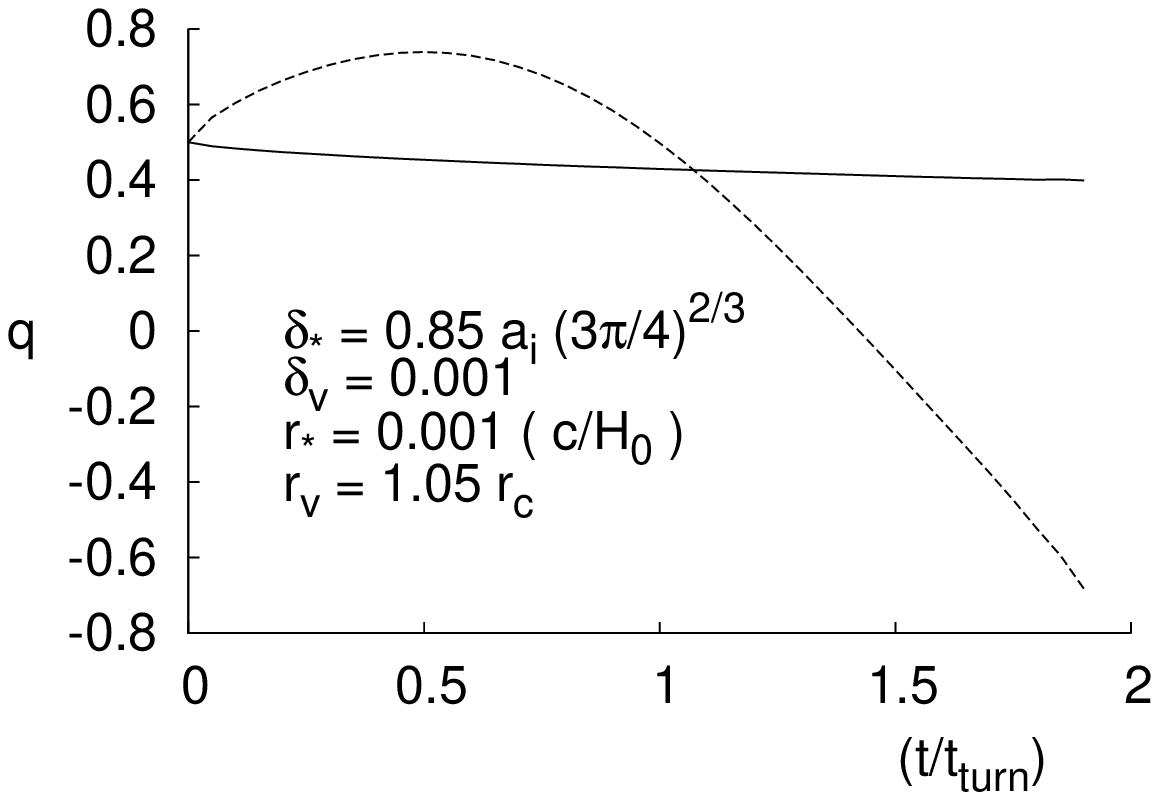}  \label{fig2a}}  
}
\fbox{
\subfloat[$t_{turn}= 1.0018 t_0$]{\includegraphics[width=.4\textwidth] 
{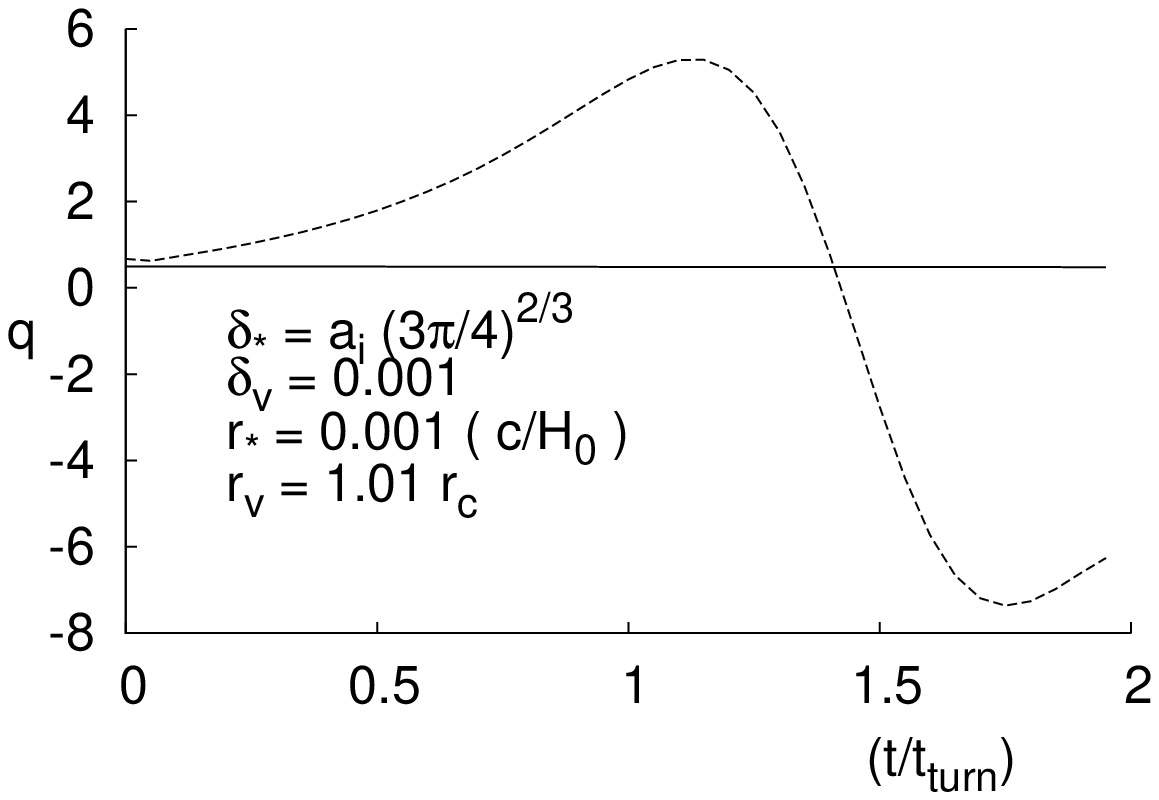}  \label{fig2b}}
}\\
\fbox{
\subfloat[$t_{turn}= 0.868 t_0$]{\includegraphics[width=.4\textwidth]
{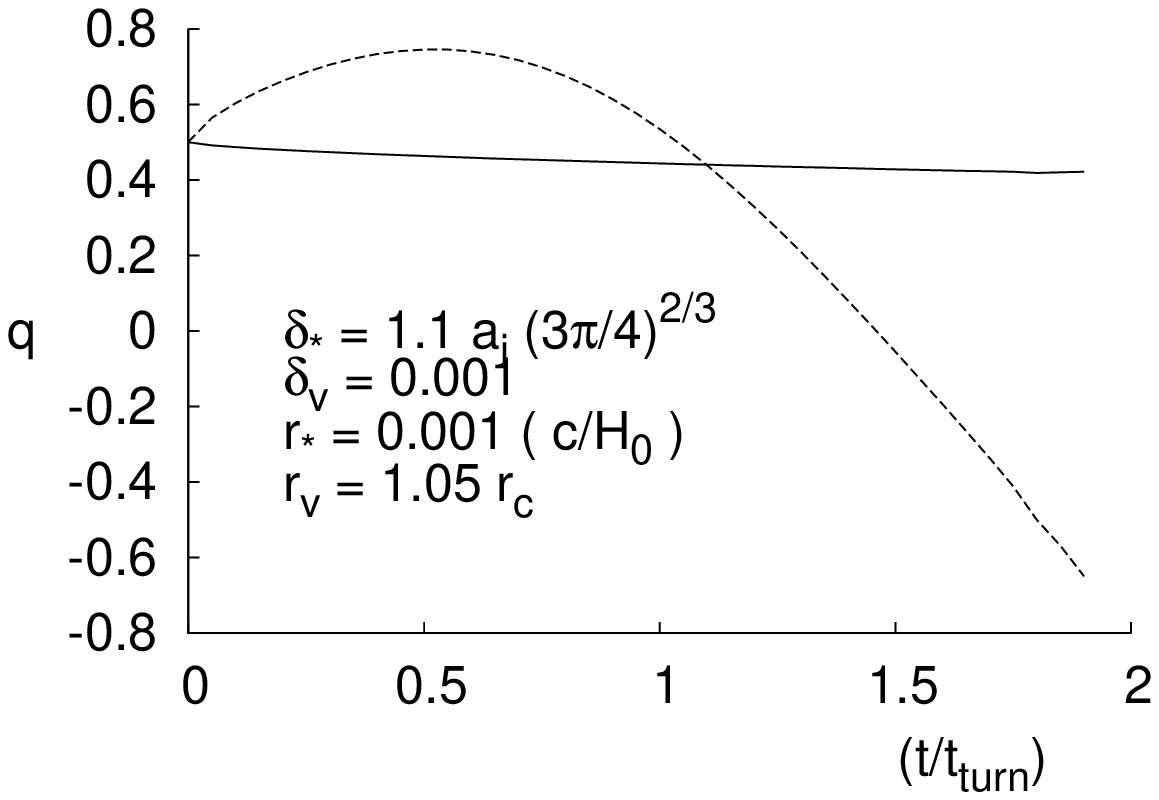}  \label{fig2c}}
}
\fbox{
\subfloat[$t_{turn}= 0.717 t_0$]{\includegraphics[width=.4\textwidth]  
{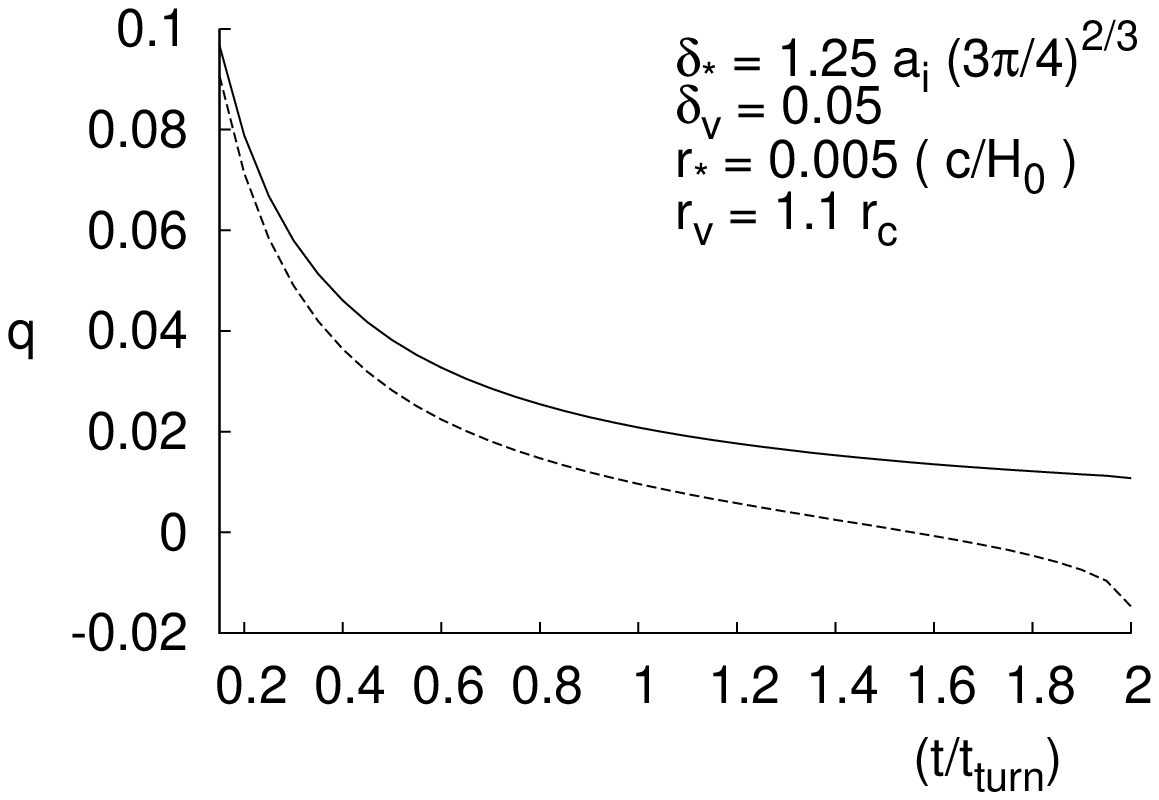}  \label{fig2d}}
}
\caption{\small The deceleration parameters for a range of parameter
  values. The dashed lines correspond to $q_{mod}$ and the solid lines
  to $q$. The $x$-axis shows $t/t_{turn}$, where $t_{turn}$ is the
  time at which Region 1 turns around, and is different for each
  plot. The values for $a_i$, $H_0$ and $c$ are the same as those
  listed in \tab{tab1}. [Both curves in \fig{fig2d} begin at
    $q\sim0.5$ at $t=t_i$.]}    
\label{fig2}
\end{figure}

For clarity, we define the volume of each of our three comoving
regions separately, as
\begin{equation}
V_1 \equiv
4\pi\int_0^{r_\ast}{\frac{R^2R^\prime}{\sqrt{1-k(r)r^2}}dr} ~~;~~ V_2
\equiv
4\pi\int_{r_\ast}^{r_c}{\frac{R^2R^\prime}{\sqrt{1-k(r)r^2}}dr} ~~;~~ 
V_3 \equiv
4\pi\int_{r_c}^{r_v}{\frac{R^2R^\prime}{\sqrt{1-k(r)r^2}}dr} \,.
\label{2eq25}
\end{equation}
The total volume of the region is used to define a volume averaged
scale factor as
\begin{equation}
a(t)\equiv \left(\frac{V(t)}{V(t_0)}\right)^{1/3} ~~;~~ V(t) \equiv 
V_1(t)+V_2(t)+V_3(t) \,,
\label{2eq26}
\end{equation}
and hence an effective deceleration parameter $q$ given by
\begin{equation}
q\equiv -\frac{\ddot a a}{{\dot a}^2} = 2 - 3\frac{\ddot V V}{{\dot
    V}^2} \,.
\label{2eq27}
\end{equation}
On the other hand, we note that R\"as\"anen's model can be mimicked
more closely by ours, if we simply remove the Region 2, by hand. By
doing so we are left with two disjoint regions, each spherically
symmetric, one of which is collapsing and the other expanding ever
rapidly and becoming ever emptier. There is no physical
reason to throw away Region 2 in this manner, but for the sake of
comparison we will define a ``modified'' scale factor $a_{mod}$ and
it's corresponding deceleration parameter $q_{mod}$ by
\begin{equation}
a_{mod}(t) \equiv \left( \frac{V_1(t)+V_3(t)}{V_1(t_0)+V_3(t_0)}
\right)^{1/3} ~~;~~  
q_{mod} \equiv -\frac{\ddot a_{mod} a_{mod}}{{\dot a_{mod}}^2}\,.
\label{2eq28}
\end{equation}
Note that the normalisation of the scale factor is irrelevant in
defining the deceleration parameter. In \fig{fig2} we plot $q(t)$ and
$q_{mod}(t)$, for several sets of initial conditions which are close
to our ``base set'' listed in \tab{tab1} (except for \fig{fig2d} which
has a large value for $\delta_v$)\footnote{Both curves in \fig{fig2d}
  begin at $q\sim0.5$ at $t=t_i$. To enhance the contrast between the
  curves, we have plotted them for times $t>0.15 t_{turn}$. The
  remaining plots (Figs. \ref{fig2a}--\ref{fig2c}) are plotted
  starting from $t=t_i$.}.  The various integrals involved in
computing $V(t)$, etc. were performed using the {\tt NIntegrate}  
routine of {\sl Mathematica}. To generate the plots we used the
numerical derivative routine {\tt ND} of {\sl Mathematica}. The
various initial conditions correspond to turnaround times that are
slightly greater than, or slightly less than, or significantly less
than the present epoch. The idea here is to demonstrate that the
results are valid regardless of whether the collapse has just begun or
is well under way at the present epoch.  
We see that while the modified scale factor does accelerate as in
R\"as\"anen's model, the scale factor $a(t)$ which is the more natural
choice in our situation, does not show this effect. The reason for
this can be understood as follows. The Region 2 is of a rather
peculiar nature -- it is underdense initially and becomes emptier with
time, however its evolution is closely linked to that of the
\emph{overdense} Region 1. Namely, the whole of Region 2 (except its
boundary at $r=r_c$), is dragged along with Region 1 and eventually
turns around, instead of expanding away to infinity like its
counterpart Region 3. Now, if one ignores Region 2, then R\"as\"anen's
arguments about the remaining two regions stand -- one region is
contracting and the other is expanding faster than the global mean,
and this stand-off leads to an acceleration of the effective scale
factor $a_{mod}$, as we see in the plots of \fig{fig2}. But if we
account for Region 2 as well, then we bring in a counter-balancing
influence of a large \emph{underdense} volume which is expanding
\emph{slower} than average, and this reduces the accelerating
influence of Region 3 to the point of making the effect completely
disappear. Note that at late times, the volume of Region 1
contributes negligibly to the total volume, and the volumes of regions
2 and 3 are comparable.

The figures \eqref{fig2a}-\eqref{fig2c} show that while the full
deceleration parameter $q$ does not change sign, it does deviate from
the EdS value of $(1/2)$, by an amount of the order of
$\sim10\%$. This indicates that while backreaction may not be large
enough to cause the scale factor to accelerate, it may still lead to
effects which are not completely negligible.

We wish to highlight two points. First, it is very important to note
the role played by the initial conditions in this entire
excercise. The function $k(r)$ is defined in a continuous fashion once
the initial density, velocity and coordinate scaling are given, and
$k(r)$ then decides which shells will eventually collapse and which
will not. The continuity of $k(r)$ assures us that in models such as
ours, with an overdensity surrounded by an underdensity, the
underdense region \emph{will always} contain a subregion in which
$k(r)>0$. We see therefore that the existence of Region 2, is a
generic feature not restricted to our specific choice of discontinuous
initial density or vanishing initial peculiar velocities. Further, as
we see in \fig{fig2d}, it is possible to make $q$ deviate even more
significantly from the EdS value than the $\sim10\%$ effect of the
first three figures, but this requires an unnaturally high value of
$\delta_v\gtrsim0.01$ (the figure has $\delta_v=0.05$), which
contradicts CMB data.  

Secondly, one may argue about the ``naturalness'' of choosing one set
of regions over another set, in order to compute volumes. But this
itself places the physicality of the acceleration effect into question
-- if one has to judiciously choose a specific set of averaging
domains in order to obtain acceleration on average, then the effect
would appear to be an artifact of this choice rather than something
which observers would see. 

Having seen that the average behaviour of the full region $0<r<r_v$ is
close to EdS, we can ask whether one can explicitly show that the
\emph{metric} for this system is actually close to FLRW. We answer
this in the affirmative in the next section.

\section{Transforming to Perturbed FLRW form}
We ask whether the metric \eqref{2eq2} can be brought to the perturbed
FLRW form with scalar perturbations, at any arbitrary stage of the
collapse. Namely, we want a coordinate transformation
$(t,r)\to(\ti{t},\ti{r})$ such that the metric in the new coordinates
is
\begin{equation}
ds^2 = -(1+2\ti{A})d\ti{t}^2 + a^2(\ti{t})(1+2\ti{\psi})\left(
d\ti{r}^2 + \ti{r}^2d\Omega^2\right)\,,
\label{3eq1}
\end{equation}
with at least the conditions
\begin{equation}
\mid\ti{A}\mid\ll1 ~~;~~ \mid\ti{\psi}\mid\ll1\,,
\label{3eq2}
\end{equation}
being satisfied. We will ignore conditions on the derivatives of
$\ti{\psi}$ and $\ti{A}$ for now (see the end of Section 3.2). The
scale factor is the EdS solution, with $\ti{t}$ as the argument. The
coordinate $\ti{r}$ is comoving with the (fictitious) background
Hubble flow, but not with the matter itself. On physical grounds we
expect that this transformation should be possible as long as the
gravitational field is weak and matter velocity is small. We will see
below that this is exactly what happens. In the new coordinates, all
matter shells labelled by $\ti{r}$ expand with the Hubble flow, with a
superimposed peculiar velocity. 

Since we want $\ti{r}$ to be comoving with the background, the natural
choice for this coordinate would be $\ti{r}\sim R/a$, at least at
early times.  Also, we need to account for the local spatial curvature
induced by the initial conditions. As an ansatz for the coordinate
transformation therefore, we consider the equations 
\begin{subequations}
\begin{equation}
\ti{r} = \frac{R(t,r)}{a(t)}\left(1 + \xi(t,r)\right) \,, 
\label{3eq3a}
\end{equation}
\begin{equation}
\ti{t} = t + \xi^0(t,r)\,,
\label{3eq3b}
\end{equation}
\label{3eq3}
\end{subequations}
where $\xi(t,r)$ and $\xi^0(t,r)$ are expected to satisfy
\begin{equation}
{\mid\xi\mid}\, \ll\, 1~~;~~ {\mid\xi^0H\mid}\, \ll\, 1\,.
\label{3eq4}
\end{equation}
This form of the transformation keeps us close to the standard
gauge transformation of cosmological perturbation theory, while still
accounting for the deviations in the evolution from the background
FLRW, caused by structure formation. We will show that a
self-consistent transformation exists, which preserves the conditions
\eqref{3eq2} and \eqref{3eq4} for most of the evolution. We will use
the metric transformation rule given by 
\begin{equation}
\ti{g}_{ab}(\ti{x})\frac{\partial\ti{x}^a}{\partial x^i}
\frac{\partial\ti{x}^b}{\partial x^j} = g_{ij}(x)\,,
\label{3eq5}
\end{equation}
and \emph{expand to leading order} in the small functions
$\xi$, $\xi^0H$, $\ti{A}$, $\ti{\psi}$ and also $k(r)r^2$ which, as we
see from \eqn{2eq17}, remains small in the entire region of
interest. The relations in \eqn{3eq5} must be analysed for the cases
$(ij)=\{(tt), (tr), (rr), (\theta\theta) \}$, in each of the three
regions. (The remaining cases can be shown to lead to trivial or
non-independent relations.) The analysis is similar to the standard
gauge transformation analysis in relativistic perturbation theory
\cite{dodel}. Since the calculations involved are straightforward but
tedious, we will only present an outline of the calculation and
highlight certain issues. At the end we will present equations for all
three regions and numerically show that the transformation is
well-behaved in the regime of interest.  

The cases $(ij)= (\theta\theta)$ and $(rr)$ are easily analysed and
lead to
\begin{equation}
 \ti{\psi} = -\xi^0H - \xi \,,
\label{3eq6}
\end{equation}
and
\begin{equation}
\xi^\prime = \frac{1}{2}k(r)r^2 \left(\frac{R^\prime}{R}\right) \,. 
\label{3eq7}
\end{equation}
The cases $(ij)= (tr)$ and $(tt)$ both require
$\mid\partial_t\ti{r}\mid\ll1$ for consistency (since the
RHS of \eqn{3eq5} in these cases has no zero order term to balance a
large $\partial_t\ti{r}$). Note that since $t$ is the proper time of
each matter shell, the quantity $\partial_t\ti{r}$ is simply the
velocity of matter in the $(\ti{t},\ti{r})$ frame (which is comoving
with the Hubble flow). In other words,   
\begin{equation}
\ti{v} \equiv \frac{\partial\ti{r}}{\partial t}\,,
\label{3eq8}
\end{equation}
is the radial comoving peculiar velocity of the matter shells in the 
$(\ti{t},\ti{r})$ frame. We will soon see that whereas the quantities
$\xi$ and $\xi^0$ behave roughly as $\sim(H_0r)^2$, the peculiar velocity
$a\ti{v}$ behaves roughly as $\sim(H_0r)$. We will therefore treat
$(a\ti{v})^2$ as a small quantity of the same order as $\xi$, etc. The
case $(ij)=(tr)$ then leads to 
\begin{equation}
\xi^{0\prime} = a\ti{v}R^\prime\,,
\label{3eq9}
\end{equation}
and the case $(ij)=(tt)$ gives
\begin{equation}
\ti{A} = -\dot\xi^0 + \frac{1}{2}(a\ti{v})^2\,.
\label{3eq10}
\end{equation}
The equations \eqref{3eq6}, \eqref{3eq7}, \eqref{3eq9}, and
\eqref{3eq10} are valid in the entire range $0<r<r_v$, provided the
peculiar velocity remains small in magnitude. The comoving peculiar
velocity is given by
\begin{equation}
\ti{v} = \partial_t\left(\frac{R}{a}\right)\,, 
\label{3eq11}
\end{equation}
where we have assumed for consistency that $|\partial_t(R/a)|\ll1$ and
have dropped the term $(R/a)\dot\xi$ since it is expected to be of
higher order than $\partial_t(R/a)$. (This can be seen from simple
dimensional considerations -- we have $\partial_t(R/a)\sim HR/a$, and
since, from \eqns{3eq7} and \eqref{2eq17}, $\xi\sim (HR)^2$, we also
have $(R/a)\dot\xi\sim (HR)^3/a$.) We will see that these conditions
do indeed hold for most of the evolution, throughout the region of
interest. 
\subsection{The transformation in Region 1 }
Since Region 1 corresponds to a homogeneous solution, the integrals in
\eqns{3eq7} and \eqref{3eq9} can be analytically performed. \eqn{3eq7}
directly gives
\begin{equation}
\xi = \frac{1}{4}k(r)r^2  =
\frac{1}{4}\left(\frac{\delta_\ast}{a_i}\right) (H_0r)^2\,, 
\label{3eq12}
\end{equation}
after setting an arbitrary function of time to zero. Hence
$\ti{v}$ has the structure $\ti{v} = r y_1(t)$, since $R$ has the
structure $R=ry_2(t)$. \eqn{3eq9} then leads to  
\begin{equation}
\xi^0 = \frac{1}{2}\ti{v}aR\,,
\label{3eq13}
\end{equation}
after setting another arbitrary function of time to zero. [Note that
  it might be more meaningful to fix the two arbitrary functions of
  time $\xi(t,0)$ and $\xi^0(t,0)$, by requiring that $\xi(t,r_c)$ and
  $\xi^0(t,r_c)$ vanish. This would be in line with the shell $r=r_c$
  expanding like the flat EdS background. However, this complicates
  some of the expressions we evaluate, and does not change the order
  of magnitude of any of the final results. Hence we will continue to
  assume that the transformation functions $\xi$ and $\xi^0$ vanish at
  $r=0$ rather than at $r=r_c$. See also the end of Section 3.2.] 
\begin{figure}[t]
\centering
\fbox{
\includegraphics[height=0.25\textheight]{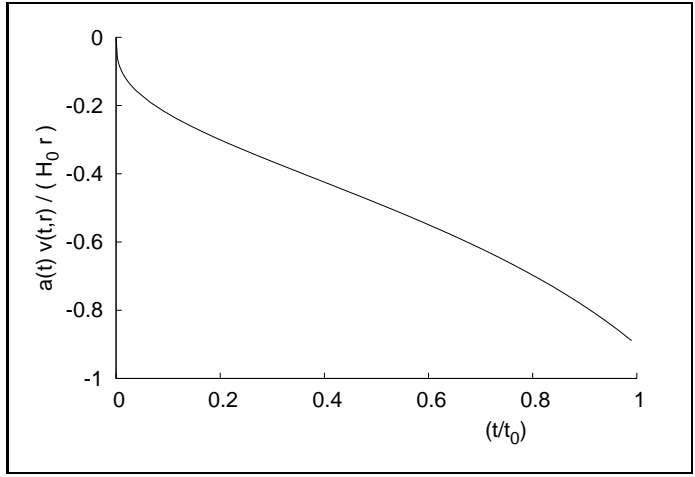}
}
\caption{\small The quantity $a\ti{v}/(H_0r)$ in Region 1, plotted
  using parameter values from \tab{tab1}. Since $(H_0r_\ast/c) \sim
  0.001$, the peculiar velocity $a\ti{v}$ remains small.}
\label{fig3}
\end{figure}

The peculiar velocity can be explicitly calculated to be
\begin{equation}
a(t)\ti{v}(t,r) = (H_0r)
\left(\frac{\delta_\ast}{a_i}\right)^{1/2}
\left[\frac{\sin{u}}{(1-\cos{u})} - \frac{2}{3}
\frac{1-\cos{u}}{(u-\sin{u}+B)} \right] \,, 
\label{3eq14}
\end{equation}
where the various functions are defined in \eqns{2eq18}, and we have
defined the constant $B$ by
\begin{equation}
B \equiv \frac{2H_0t_i}{1+\delta_\ast} \left( \frac{\delta_\ast}{a_i}
\right)^{3/2} - \left(u_i-\sin{u_i} \right)\,.
\label{3eq15}
\end{equation} 
In the rest of this section we will use the parameter values listed in 
\tab{tab1}. In \fig{fig3} we have plotted $a\ti{v}/(H_0r)$ in Region
1. We see that this dimensionless quantity remains of order $\sim1$
throughout the evolution. For our choice of
$(H_0r_\ast/c)\sim10^{-3}$, which corresponds to an overdensity
spanning a few Mpc today, the peculiar velocity is of order
$\sim10^{-3}$ in Region 1.  

\begin{figure}[t]
\centering
\fbox{
\subfloat[]{
  \includegraphics[width=0.4\textwidth]{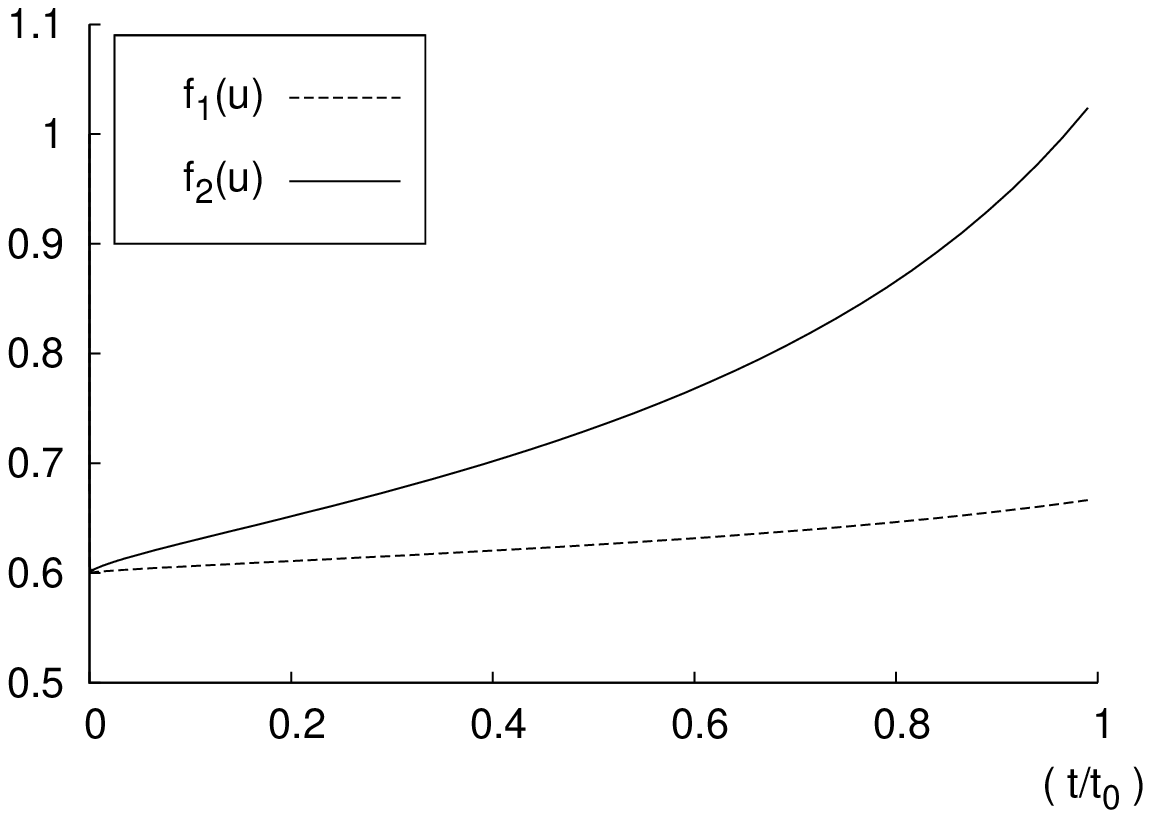}
\label{fig4a} }  
}
\fbox{
\subfloat[]{
  \includegraphics[width=0.4\textwidth]{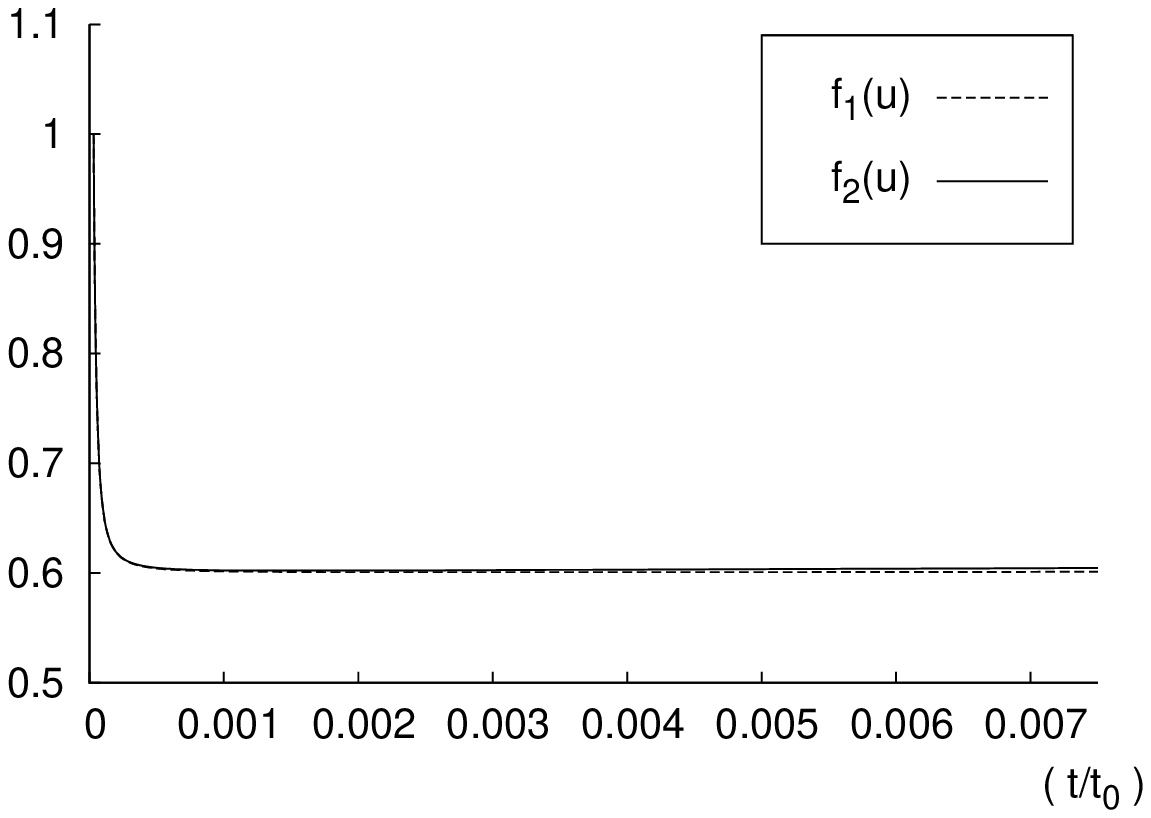}
\label{fig4b} }
}
\caption{\small The time dependence of the functions $\ti{\psi}$ and
  $\ti{A}$ in Region 1, using parameter values from \tab{tab1}. Panel
  (a) shows the evolution from $t_i$ to $t_0$, and panel (b) shows the
  early time evolution upto $t=0.0075t_0$. The dotted lines correspond
  to $f_1(u(t))$ and the solid lines to $f_2(u(t))$ (see \eqns{3eq16}
  and \eqref{3eq17} for definitions). We see that at early
  times, $\ti{A}\simeq-\ti{\psi}$. Since $(H_0r_\ast/c)\sim0.001$,
  $\ti{\psi}$ and $\ti{A}$ remain small.}   
\label{fig4}
\end{figure}

Knowing $\xi$ and $\xi^0$ we can easily determine $\ti{\psi}$ and
$\ti{A}$ to be
\begin{align}
\ti{\psi} &= -\frac{1}{4}\left(\frac{\delta_\ast}{a_i}\right) (H_0r)^2
- \frac{1}{2} a\ti{v}R H \nonumber\\
&= -\frac{1}{4}\left(\frac{\delta_\ast}{a_i}\right) (H_0r)^2 \left[1 -
  \frac{4}{3}\left(\frac{2}{3}\frac{(1-\cos{u})^2}{(u-\sin{u}+B)^2} -
  \frac{\sin{u}}{(u-\sin{u}+B)} \right) 
  \right]\,, \nonumber\\
&\equiv -\frac{1}{4}\left(\frac{\delta_\ast}{a_i}\right) (H_0r)^2
f_1(u)\,, 
\label{3eq16}\\
& \nonumber\\
\ti{A} &= \frac{1}{4}\left(\frac{\delta_\ast}{a_i}\right) (H_0r)^2
\left[ \frac{4}{3} \left(2\frac{\sin{u}}{u-\sin{u}+B} -
  \frac{(1-\cos{u})^2}{(u-\sin{u}+B)^2} \right) -
  2\frac{\cos{u}}{1-\cos{u}}  \right] + \frac{1}{2}(a\ti{v})^2 \,,
\nonumber\\ 
& \equiv \frac{1}{4}\left(\frac{\delta_\ast}{a_i}\right) (H_0r)^2
f_2(u)\,, 
\label{3eq17}
\end{align}
where $a\ti{v}$ is given in \eqn{3eq14}, and the last equalities in
the equations define the functions $f_1(u(t))$ and $f_2(u(t))$
respectively. In \fig{fig4} we have plotted the functions $f_1(u(t))$
and $f_2(u(t))$. We see that they remain of order $\sim1$ for most of
the evolution, and hence for an overdensity spanning a few Mpc,
$\ti{\psi}$ and $\ti{A}$ are of order $\sim10^{-6}$ in Region 1. By
expanding $f_1(u)$ and $f_2(u)$ in the parameter $u$ around its
initial value $u_i$, one can show that at early times one has
$\ti{A}\simeq\,-\ti{\psi}$, as expected in the linear
theory. \fig{fig4b} shows this behaviour. However, the relative
difference between $\ti{A}$ and $-\ti{\psi}$ grows quickly and (for
the parameters given in \tab{tab1}) becomes of the order $10^{-2}$, by
$t\sim 0.02t_0$. At the end of Section 3.2 we show that the large
relative difference between $\ti{A}$ and $-\ti{\psi}$ (of the order
$\sim50\%$ at late times), is largely due to our choice of setting
these functions to zero at the origin. [We note that a difference
between these two functions is not in principle surprising since the
density contrast in Region 1 grows to order $\sim1$ by $t\sim0.3t_0$
(see \fig{fig1a}), and departures from the linear theory are expected
to become significant even before this. See, however, Section 4
below. ]  

\subsection{The transformation in Regions 2 and 3}
The calculation in Regions 2 and 3 proceeds in a similar fashion as
above, but in this case the integrals involved cannot be computed
analytically. We will therefore display the expressions we obtain for
$\ti{v}$, $\xi$,  and $\xi^0$, and plot the results of numerically
computing $\ti{\psi}$ and $\ti{A}$ from these quantities.  
\begin{itemize}
\item {\bf Region 2 ($r_\ast<r<r_c$):} \\
For numerical calculations we found it convenient to define the
functions 
\begin{subequations}
\begin{align}
B_1(r) &\equiv \frac{1}{2}\frac{H_0^2}{a_i} r \left( \varepsilon -
  \frac{r\varepsilon^\prime}{1+\varepsilon} \right)\,,
\label{3eq18a}\\
B_2(t,r) &\equiv \frac{1}{2}\frac{H_0^2}{a_i}
\frac{r^2\varepsilon^\prime}{(1+\varepsilon)}
\frac{\varepsilon^{3/2}}{(1-\cos{\alpha})^2} \left[H_i(t-t_i)\sin{\alpha}
  \frac{(3+\varepsilon)}{(1+\varepsilon)} 
+  \frac{4\varepsilon^{1/2}}{(1+\varepsilon)^2}
  \frac{\sin{\alpha}}{\sin{\alpha_i}}\right] \,,
\label{3eq18b}\\
B_3(t,r) &\equiv \frac{H_i\sin{\alpha}}{(1-\cos{\alpha})^2}
\frac{2\varepsilon^{3/2}}{1+\varepsilon}\,.
\label{3eq18c}
\end{align}
\label{3eq18}
\end{subequations}
We then have
\begin{align}
&\xi(t,r) = \frac{1}{4} \left(\frac{\delta_\ast}{a_i}\right)
  (H_0r_\ast)^2 + \int_{r_\ast}^r{\left(B_1(\bar r)+B_2(t,\bar
    r)\right) d\bar r}  \,,
\label{3eq19}\\
&\ti{v} = \frac{R}{a} \left[ B_3(t,r) - H 
\right] \,,
\label{3eq20}\\
&\xi^0(t,r) = \xi^0(t,r_\ast) + a(t)\int_{r_\ast}^r{ \ti{v}(t,\bar r) 
  R^\prime(t,\bar r)d\bar r}\,,
\label{3eq21}
\end{align}
where $\xi^0(t,r_\ast)$ is computed from \eqn{3eq13} at
$r=r_\ast$. $\ti{\psi}$ and $\ti{A}$ must now be computed using
\eqns{3eq6} and \eqref{3eq10} respectively. We have again used the
     {\tt NIntegrate} routine of {\sl Mathematica}.  

\begin{figure}[t]
\centering
\fbox{
\includegraphics[height=0.25\textheight]{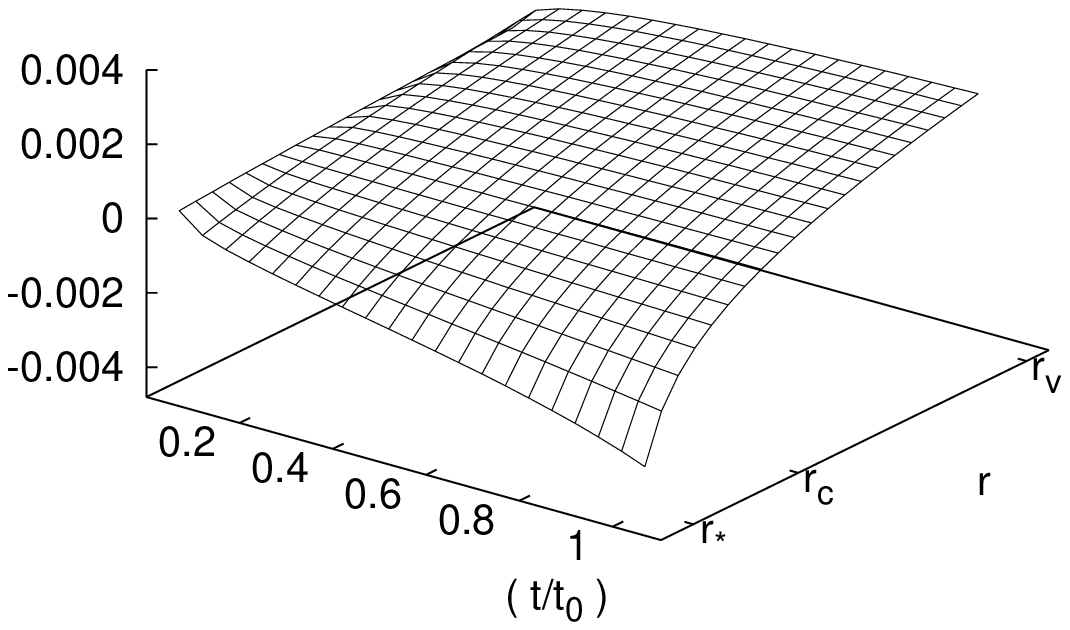}
}
\caption{\small The peculiar velocity $a\ti{v}/c$ in
  Regions 2 and 3 using parameter values from \tab{tab1}.}
\label{fig5}
\end{figure}
\item {\bf Region 3 ($r_c<r<r_v$) :}\\
The analysis is very similar to that in Region 2. We define the
functions
\begin{subequations}
\begin{align}
D_1(r) &\equiv \frac{1}{2}\frac{H_0^2}{a_i} r \left( \varepsilon -
  \frac{r\varepsilon^\prime}{1+\varepsilon} \right)\,,
\label{3eq22a}\\
D_2(t,r) &\equiv \frac{1}{2}\frac{H_0^2}{a_i}
\frac{r^2\varepsilon^\prime}{(1+\varepsilon)}
\frac{|\varepsilon|^{3/2}}{(\cosh{\eta}-1)^2} \left[H_i(t-t_i)\sinh{\eta}  
  \frac{(3+\varepsilon)}{(1+\varepsilon)} 
+  \frac{4|\varepsilon|^{1/2}}{(1+\varepsilon)^2}
  \frac{\sinh{\eta}}{\sinh{\eta_i}}\right] \,,
\label{3eq22b}\\
D_3(t,r) &\equiv \frac{H_i\sinh{\eta}}{(\cosh{\eta}-1)^2}
\frac{2|\varepsilon|^{3/2}}{1+\varepsilon}\,,
\label{3eq22c}
\end{align}
\label{3eq22}
\end{subequations}
and find
\begin{align}
&\xi(t,r) = \xi(t,r_c) + \int_{r_c}^r{\left(D_1(\bar r)+D_2(t,\bar
    r)\right) d\bar r}  \,,
\label{3eq23}\\
&\ti{v} = \frac{R}{a} \left[ D_3(t,r) - H 
\right] \,,
\label{3eq24}\\
&\xi^0(t,r) = \xi^0(t,r_c) + a(t)\int_{r_c}^r{ \ti{v}(t,\bar r) 
  R^\prime(t,\bar r)d\bar r}\,,
\label{3eq25}
\end{align}
where $\xi(t,r_c)$ and $\xi^0(t,r_c)$ are obtained from \eqns{3eq19}
and \eqref{3eq21} respectively, evaluated in the limit $r\to r_c^-$.
\end{itemize}

In \fig{fig5}, we have plotted the velocity $a\ti{v}/c$ in Regions 2
and 3 for a range of time. It can be shown that at the order of
approximation we are working at, $a\ti{v}$ changes sign
at $r=r_c$. [ Recall $\varepsilon(r_c)=0$ and hence this shell expands
  exactly like the EdS background. The metric in the $(\ti{t},\ti{r})$
  coordinates will not be exactly EdS at $r=r_c$, due to our unusual
  choice of normalisation for $\xi$ and $\xi^0$ at $r=0$. This does
  not pose any problem for our conclusions.]  

\begin{figure}[t]
\centering
\fbox{
\subfloat[$\ti{\psi}(t,r)$]{
  \includegraphics[width=0.4\textwidth]{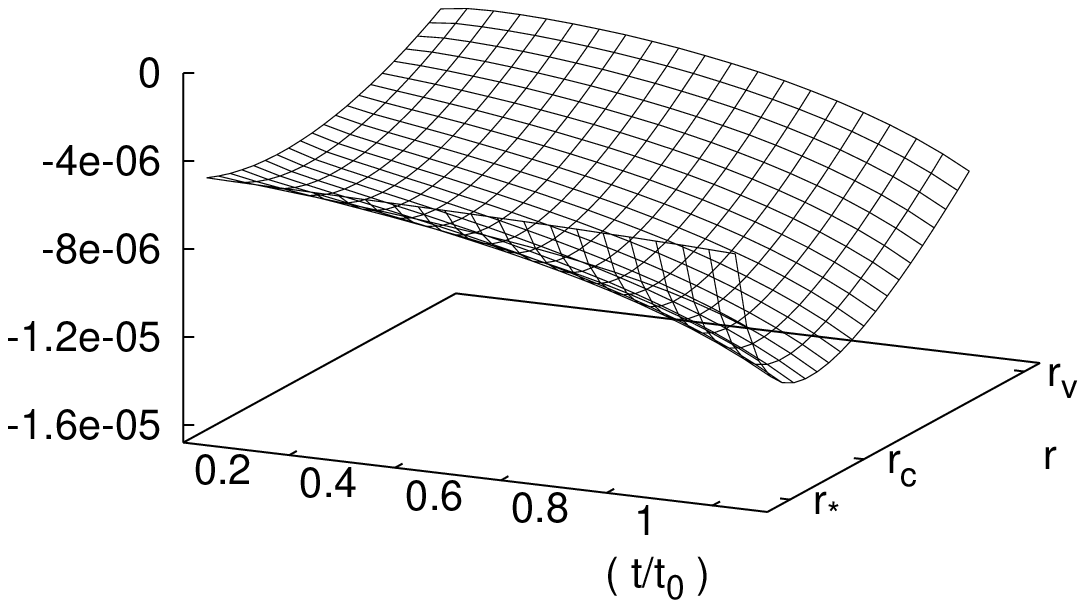} \label{fig6a}}
}
\fbox{
\subfloat[$\ti{A}(t,r)$]{
  \includegraphics[width=0.4\textwidth]{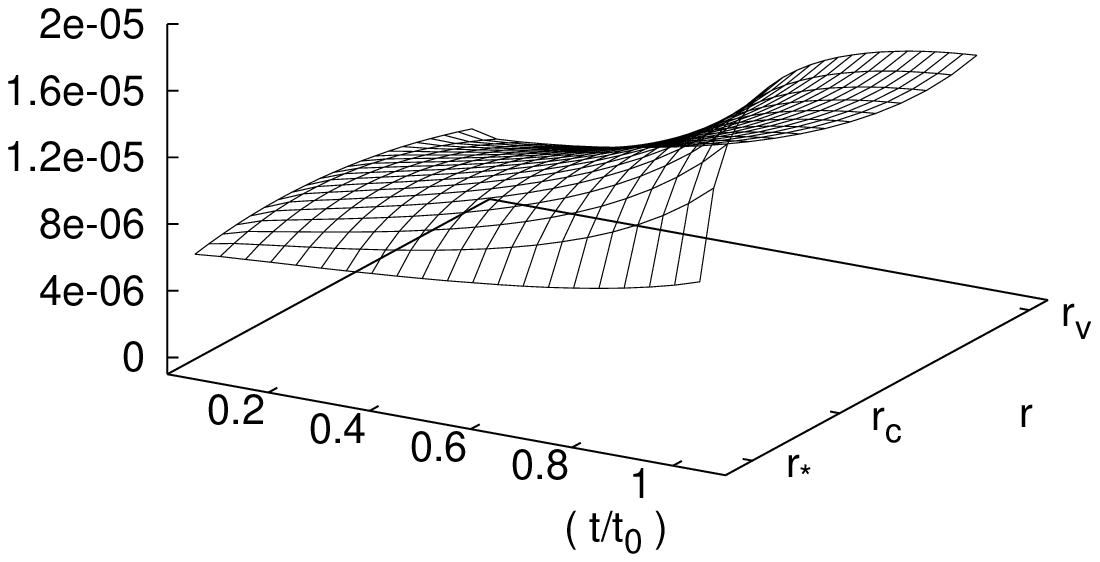} \label{fig6b}}
}
\caption{\small The metric functions $\ti{\psi}(t,r)$ and
  $\ti{A}(t,r)$ in Regions 2 and 3 using parameter values from
  \tab{tab1}. The time axis begins at $t=50t_i$.}   
\label{fig6}
\end{figure}

In \fig{fig6} we plot $\ti{\psi}$ and $\ti{A}$.
We see that these functions are well behaved and remain small for the
entire region of interest (in space and time). Hence the perturbed
FLRW picture is indeed valid for this system, even though each region
by itself appears to be very different from FLRW in the synchronous
coordinates comoving with the matter. Due to numerical difficulties
close to the initial time $t=t_i$, we have plotted the time axis
starting from $t=50t_i$. 

Note that the magnitude of $\ti{\psi}$ and $\ti{A}$ is sensitive to
the overall size of the region, determined by the value of
$R(t,r_v)$. For our parameter choices given in \tab{tab1}, the size of
the region at the present epoch is $\sim33 {\rm Mpc}$, which is a
typical size for observed voids. The dependence is roughly $(HR)^2$,
and hence a void which is about 10 times larger in length scale than
the above value, would have metric functions about 100 times larger.

\begin{figure}[t]
\centering
\fbox{
\subfloat[$r=(1/2)(r_\ast+r_c)$]{
  \includegraphics[width=0.4\textwidth]{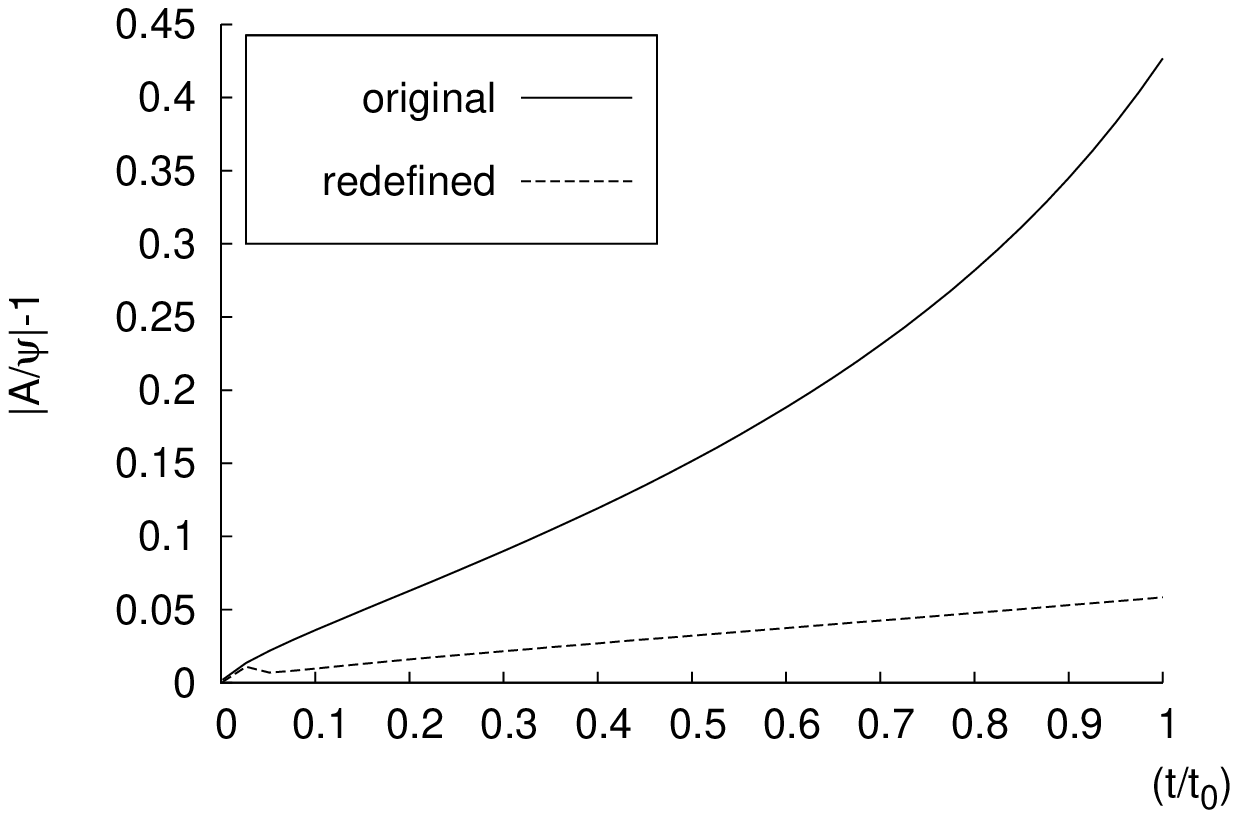} \label{fig7a}}
}
\fbox{
\subfloat[$r=(1/2)(r_c+r_v)$]{
  \includegraphics[width=0.4\textwidth]{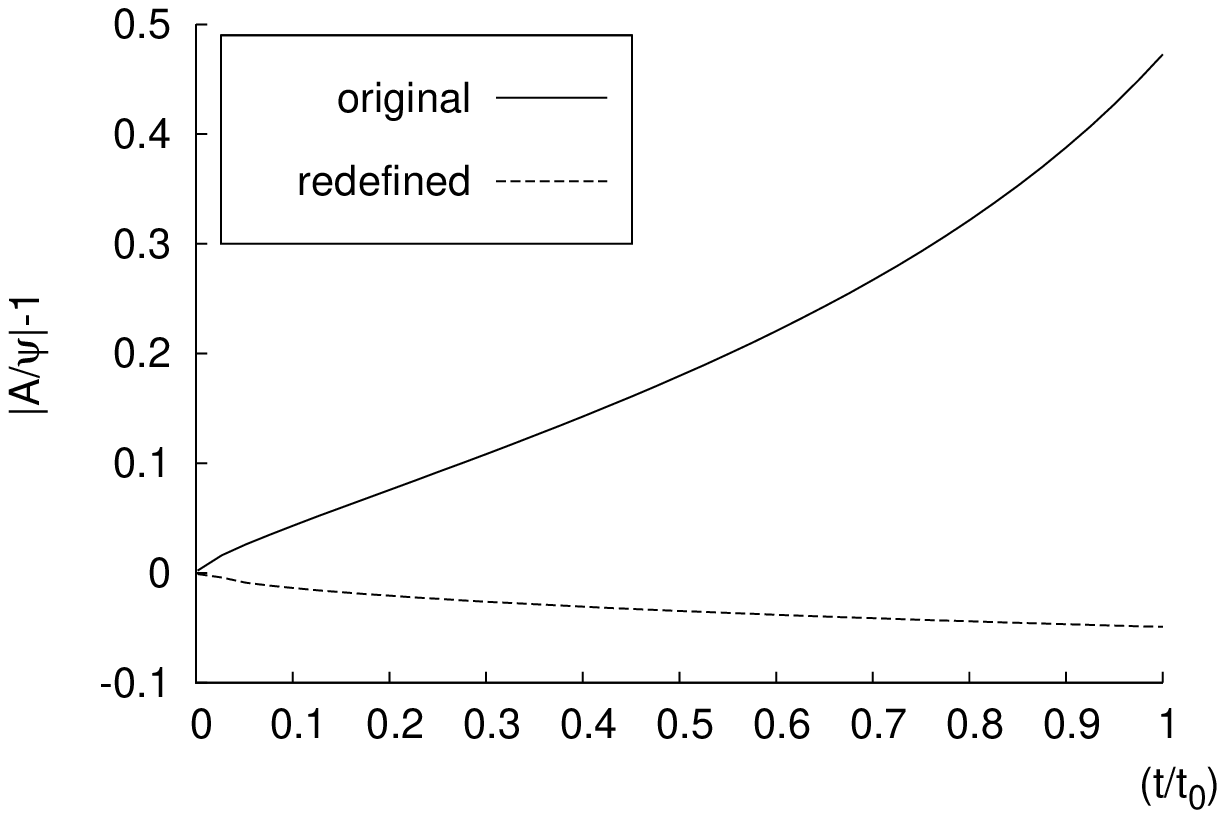} \label{fig7b}}
}
\caption{\small The relative difference between the metric functions
  $\ti{A}$ and $-\ti{\psi}$, defined as $|\ti{A}/\ti{\psi}|-1$,
  plotted as a function of time at $r=(1/2)(r_\ast+r_c)$ (panel (a))
  and at $r=(1/2)(r_c+r_v)$ (panel (b)). The solid lines show the
  difference for the definition of the functions used in this
  paper. The dotted lines show the difference when the functions are
  redefined to vanish on $r=r_c$ instead of vanishing at the origin.}   
\label{fig7}
\end{figure}

We end this subsection by noting two points. Firstly, we have seen
that with our choices for certain arbitrary functions of time in
\eqns{3eq12} and \eqref{3eq13}, the functions $\ti{A}$ and
$-\ti{\psi}$ build up a relative difference of $\sim40$-$50\%$ by
present epoch (see Figs. \eqref{fig4a} and \eqref{fig6}). However, it
can be shown that if one redefines these functions so that they vanish
at $r=r_c$ rather than at the origin (in line with the discussion
below \eqn{3eq13}), then the relative difference between the functions
reduces to $\sim5\%$ at present epoch in Regions 2 and 3. This is
demonstrated in \fig{fig7} for two representative values of the
coordinate $r$. In Region 1 the the relative difference is maximum at
the center $r=0$, being about $30\%$ at present epoch (which is
obvious since with the redefined functions, the relative difference at
the center is simply the relative difference of the \emph{original}
functions at $r=r_c$). The origin of the large relative difference
seen in, e.g. \fig{fig4a} is therefore not completely clear, and may
even be unphysical. This issue deserves more careful consideration,
and we hope to return to it in future work (see also Section 4
below). In any case, we emphasize that our main result is that the
magnitude of the functions $\ti{A}$ and $\ti{\psi}$ themselves, is
very small. This brings us to the second issue. 

It is known that simply having a metric of the form \eqref{3eq1} with
only the \emph{magnitude} of the perturbations being small, is not
enough to guarantee consistency with Einstein's equations written as a
perturbation series; additional constraints on the \emph{derivatives}
of these functions must be satisfied. These constraints, given in
e.g. \Cite{wald}, take the form (for the metric \eqref{3eq1} with
$\ti{\psi}=-\ti{A}$),
\begin{equation}
\left| \frac{\partial\ti{A}}{\partial t} \right|^2 \ll
\frac{1}{a^2} \nabla^\alpha\ti{A} \nabla_\alpha\ti{A} \,,~~~
\left(\nabla^\alpha\ti{A} \nabla_\alpha\ti{A} \right)^2 \ll
\left(\nabla^\alpha\nabla^\beta\ti{A} \right)
\nabla_\alpha\nabla_\beta\ti{A}\,, 
\label{3eq26} 
\end{equation}
where $\alpha,\beta=1,2,3$, and $\nabla_\alpha$ is the spatial
covariant derivative associated with the flat 3-space metric. On
dimensional grounds, treating $\ti{A}\sim (HR)^2 \ll1$, $\partial_t
\sim H$ and $\nabla\sim aR^{-1}$, it is easy to see that these
constraints will be satisfied by our solution. This should also be
expected since we started from an exact solution of the Einstein
equations and performed a self-consistent coordinate transformation. 

\subsection{The magnitude of the backreaction}
One can now legitimately ask the question, ``How large is the effect
of the small metric inhomogeneities?'' Naively, one would argue that
small inhomogeneities must lead to small effects. Indeed, the question
of the magnitude of the backreaction in the Newtonianly perturbed FLRW
setting has been investigated by Behrend, \emph{et al.} \cite{behr} in
the linear and quasilinear regimes, and they find that corrections to
the FLRW equations remain at the level of one part in $10^5$. The
effect of perturbative metric inhomogeneities on observables such as
the luminosity distance \emph{evaluated using the perturbed FLRW
  metric with a fixed background}, has also been studied \cite{flan}
and shown to be small. However, what we are dealing with is a
situation in which the \emph{matter} perturbations are completely
nonlinear, and it is not \emph{a priori} clear that the same arguments
would carry through. Indeed, we saw in section 2 that the deceleration
parameter $q$ deviated from its EdS value by about $\sim10\%$. Here we
give an argument based on dimensional considerations supplemented with
realistic numbers, which will show that this effect is scale
dependent, and is not expected to be present if a sufficiently large
averaging scale is chosen.

In the following we will work at the present epoch $t_0$. Consider a
model situation similar to the one we have been considering so far,
such that at present epoch the physical extent of the overdense region
is $R_\ast$, and that of the underdense is $R_v$. For order of
magnitude estimates, we assume that in the perturbed FLRW metric
\eqref{3eq1} (which is valid for this system provided $H_0R_v \ll1$),
$\ti{A} \sim -\ti{\psi}$. Also assume that the density contrast in the
overdense region is $\delta_{\ast0}$ and that in the underdense region
is $\delta_{v0}$, where we take $\delta_{\ast0}$ and $\delta_{v0}$ to
be constant in space, which is fine for an order of magnitude
estimate. The backreaction in the Buchert approach contains, among
other terms, the spatial average of the quantity $\nabla^2\ti{A}$
which appears in the spatial curvature \cite{buch,behr}, where
$\nabla^2$ is the Laplacian operator for the flat $3$-space
metric. The spatial curvature has the structure 
\begin{equation}
\mathcal{R} \sim \frac{1}{a^2}\left[(\#_1) \nabla^2\ti{A} +
  (\#_2) \ti{A}\nabla^2\ti{A} + (\#_3) (\nabla\ti{A})^2 \right]\,,
\label{3eq27}
\end{equation}
where $\#_1,\#_2,\#_3$ are constants whose values are irrelevant for
this order of magnitude argument. Due to the Einstein equations in the
small scale Newtonian approximation, the leading order effect in the
\emph{nonlinear} regime, comes from $\nabla^2\ti{A}$ which satisfies
\begin{equation}
\nabla^2\ti{A} \sim\,\, \left\{
\begin{array}{l}
H_0^2\delta_{\ast0} \,,~~~~~ \text{overdense region}\,,\\
H_0^2\delta_{v0} \,,~~~~~ \text{underdense region}\,.
\end{array} \right .
\label{3eq28}
\end{equation}
Consider the situation when, at present epoch,
$R_\ast \sim 6$Mpc, $R_v \sim 30$Mpc, $\delta_{\ast0} \sim 10^2$
and $\delta_{v0} \sim -0.9$. These are typical numbers for clusters
of galaxies and voids. It is straightforward to now show that the
spatial average of $\nabla^2\ti{A}$ over a domain comprising the
overdense and underdense region, works out to be 
\begin{align}
 \langle \nabla^2\ti{A} \rangle &\sim \frac{H_0^2}{R_\ast^3 + R_v^3}
 \left[ R_\ast^3 \delta_{\ast0}  + R_v^3\delta_{v0} \right]\,,\nonumber \\
&\simeq -0.1 H_0^2\,.
\label{3eq29}
\end{align}
It would appear  therefore, that this spatial average of
$\nabla^2\ti{A}$ (which is usually neglected) thus turns out to be a
significant contributor to the backreaction. (In fact it is the most
significant contributor, since the other terms are clearly of at least
one higher order in the small quantity $(H_0R_v)^2$, for such a
model.) 

As we now argue, however, the above effect can be deceptive, and is
really scale dependent. Let the initial density contrasts in the
overdense and underdense regions be $\delta_{\ast i}$ and
$\delta_{vi}$ respectively, so that $\delta_{\ast
  i},|\delta_{vi}|\ll1$. If $M_{\ast i}$, $M_{vi}$, $M_{\ast}$ and 
$M_v$ are the masses at initial time and today, in the overdense and
underdense region respectively, and $\rho_i$ and $\rho_0$ are the
values of the background density at initial time and today, then at
initial time 
\begin{equation}
M_{\ast i}  \approx \rho_i (a_i R_\ast)^3 = \rho_0 R_\ast^3\,,~~
M_{vi} \approx \rho_0 R_v^3 
\,, 
\label{3eq30}
\end{equation}
and at present time,
\begin{equation}
M_\ast = \rho_0 (1 + \delta_\ast) R_\ast^3  >  M_{\ast i}\,, ~~ M_v =
\rho_0 (1 + \delta_v) R_v^3  <  M_{vi}\,.
\label{3eq31}
\end{equation}
We now make the crucial observation that \emph{if the averaging scale
  is large enough}, and we are counting several such ``pairs'' of
overdense and underdense regions, then the mass ejected from the
underdense region must have all gone into the overdense region. It is
then easy to show, that
\begin{equation}
\delta_\ast R_\ast^3 \approx - \delta_v R_v^3\,,
\label{3eq32}
\end{equation}
which means that, just like in the linear theory, the average of
$\nabla^2\ti{A}$ is expected to be negligible on such a scale. In the
real universe, we do expect that the averaging scale must be at least
of the order of the homogeneity scale, and on such a scale we will be
sampling several pairs of overdense and underdense regions. The only
cumulative effects that may arise with such a choice of scale are
from terms such as $(\nabla\ti{A})^2$, which as we mentioned earlier,
are of one higher order in the perturbation and will give effects of
the size $\sim H_0^2 (H_0R_v)^2 \ll H_0^2$. (For a demonstration of
the scale dependence of the effect, see e.g. the work of Li and
Schwarz \cite{schwarz2}.) 

In the light of this discussion, our results of section 2 (figures 
\eqref{fig2a}-\eqref{fig2c}) can be understood as arising due to the
fact that we are not averaging on a scale large enough to sample many
overdense and underdense pairs. In the language of the above argument,
all the mass ejected from the underdense region (in coordinates which
are comoving with the \emph{background}), is not going into the
overdense region which is part of the averaging domain. Some of the
matter is escaping outside the averaging domain and is left
unaccounted for. This explains an imbalance as suggested by
\eqn{3eq29}.

\section{Discussion}
In this paper we have addressed the question of whether or not the
assumptions involved in using a perturbed FLRW framework to describe
the \emph{metric} of the present Universe, break down during epochs of
structure formation, and whether averaging over inhomogeneities in
such a context can yield effects large enough to lead to an
accelerating scale factor. We have demonstrated two things. First, it
is very essential to impose initial conditions properly matched across
any boundaries present in the model. Neglecting to do so (for example
by ignoring certain regions such as Region 2 of our model) can lead to
effects such as an accelerating effective scale factor
$a_{mod}$. Also, as we saw in \fig{fig2d}, it \emph{is} in principle
possible to have very large deviations from FLRW-like conditions after
averaging, but this is at the cost of initial conditions which are
unrealistic when confronted with the CMB data.

Secondly, we have shown that, provided peculiar velocities of matter 
are small (which is a reasonable assumption on observational grounds),
one can explicitly find coordinates in which the metric of the
Universe is of the perturbed FLRW form \eqref{3eq1}. This by itself is
not a surprising result, since one can always do this locally (our
Region 1 for example is a closed FLRW sub-Universe by itself, without
any coordinate transformations). What is important is the fact that
the background FLRW used for comparison is taken to be uniquely and
\emph{globally} defined, so that \emph{every} local region involved in
structure formation can be compared to this background in an
unambiguous fashion. The existence of this global FLRW model is not
simply an assumption made at the present epoch, it is justified by the
fact that density fluctuations in the \emph{past} were very small.

An issue worth highlighting again is that the time coordinates
involved in the transformation from the synchronous coordinates to the
perturbed FLRW form, are linked by an infinitesimal transformation
(after controlling some degrees of freedom by setting arbitrary
functions of time to zero), as expected with weak gravitational
fields. While our model is perhaps not the most realistic depiction of
observed voids, we expect this feature to hold in more realistic
models also. At first glance this may appear to differ from
Wiltshire's claim \cite{wilt} that the clocks of observers in bound
regions face a significant calibration when compared with those inside
large voids. However the issue is rather subtle, since Wiltshire's
arguments involve a different choice of time slicing than the standard
constant $t$ hypersurfaces used here, and his model shows cumulative
effects over large times. It is possible that these cumulative effects
are related to the relative difference between the metric functions
which we discussed in Section 3, but a straightforward comparison is
not possible at this stage. In this context, it becomes essential
to understand the origins of those differences, and moreover the fact
that the magnitude of the difference is sensitive to boundary
conditions. 

A few remarks concerning further tests of the backreaction
argument, with more realistic models of structure formation which
should account for pressure. We have not modelled voids
\emph{surrounded} by overdensities, since such models are plagued by
shell-crossings in the absence of pressure. (Our current model also
faces this difficulty beyond region 3.) But qualitatively, we expect
our results to hold even in models where a suitable pressure term
takes care of shell crossings and leads to stable structures
surrounding expanding voids, since we expect peculiar velocities to
remain small in this situation as well. As long as the sizes of
individual voids are not an appreciable fraction of the Hubble scale,
one expects that here as well, it will be possible to find coordinate
transformations like the one we have presented. We will return to this
problem in future work.   

Finally, 
we comment on an issue which deserves careful consideration. Recall
that in the Buchert scheme of averaging perturbative inhomogeneities,
which has been employed by several authors
\cite{kolb-notari,behr,schwarz}, there are \emph{two} scale 
factors -- the background scale factor $a(t)$ which satisfies the
usual FLRW equations, and the effective scale factor $a_D(t)$ which
satisfies the Buchert equations \cite{buch}. It is our belief that
this situation has an inherent ambiguity -- which of the two scale
factors is relevant for observations? In our opinion, a more
consistent way of proceeding would be to employ the fully covariant
averaging scheme due to Zalaletdinov \cite{zala}, developed further in
the cosmological context by \cite{pasi}. In this scheme, there would
be only \emph{one} scale factor, unambiguously associated with a
background metric, which  would satisfy the \emph{corrected} Einstein
equations. The fact that backreaction is negligible
in the linear regime should allow the problem to be well-posed at say
the last scattering epoch. While we have estimated the effects of
averaging inhomogeneities to be negligible on large enough scales in
Buchert's scheme, this was done in a perturbation theory around the
\emph{fixed} Einstein-deSitter background. It is not clear that this
result will continue to hold even at late times when the background
itself is ``evolving'', namely, being corrected by the (small but
cumulative) effects of inhomogeneities. Note that \emph{if} these
effects build up at late times, then in this scheme with only one
scale factor, the evolution of the perturbations will also be
modified, which may lead to some interesting effects.
We will investigate these issues in the near future.

In conclusion, we wish to suggest that it has not yet been
conclusively established that cosmological backreaction becomes large
enough during the structure formation phase to cause an acceleration
of the scale factor. On the contrary, the calculation presented in the
present paper suggests that the backreaction remains small during this
phase, and that the intuitive picture concerning weak gravitational
fields is actually realized in the form of a perturbed FLRW
metric. Self-consistently accounting for the effects that the
backreaction would implicitly have on the evolution of the
perturbations (via the scale factor), may lead to interesting effects
which remain to be explored.

{\bf Acknowledgments :} It is a pleasure to thank Karel van Acoleyen and
T. Padmanabhan for several useful discussions. We are also grateful to
Thomas Buchert, Syksy R\"as\"anen, Dominik Schwarz and David Wiltshire
for their critical comments on an earlier draft.

\end{document}